\documentclass[a4paper,11pt]{article}
\pdfoutput=1 
\usepackage{jcappub}
\usepackage[T1]{fontenc}
\usepackage[dvipsnames]{xcolor}
\usepackage{tikz}
\usetikzlibrary{patterns}
\usetikzlibrary{patterns.meta}
\usepackage{verbatim}
\usepackage{subcaption}
\usepackage[normalem]{ulem}

\title{\boldmath Nonlinear Evolution of the Matter Trispectrum with Primordial Parity Violation}

\author[a,b,\textsuperscript{$\dagger$}]{Sha Azyzy,}
\author[a]{Drew Jamieson,}
\author[a,b,c]{Eiichiro Komatsu,}
\author[a]{and Toshiki Kurita}

\affiliation[a]{Max Planck Institute for Astrophysics, Karl-Schwarzschild-Str.1, 85741 Garching, Germany}
\affiliation[b]{Ludwig-Maximilians-Universität München, Schellingstr. 4, D-80799 München, Germany}
\affiliation[c]{Kavli Institute for the Physics and Mathematics of the Universe (Kavli IPMU, WPI),
UTIAS, The University of Tokyo, Chiba 277-8583, Japan}

\abstract{
Parity-odd four-point correlation functions, or trispectra, of cosmic matter density fields provide a unique probe of fundamental symmetries in cosmology. Trispectra of primordial matter density fluctuations produced in the early universe are modified by the subsequent nonlinear structure formation. In this paper, we compute the nonlinear evolution of the parity-odd matter trispectrum to one-loop order, i.e., to third order in density fluctuations, within the framework of effective field theory of the large-scale structure of the universe. By analyzing the different terms in the perturbation series, we demonstrate the structure of infrared divergence cancellations, as required by the equivalence principle. We also derive the forms of the counterterms required to renormalize the ultraviolet divergences. Adopting a specific model for a primordial parity-odd trispectrum, we numerically compute the leading-order effects of nonlinear gravitational evolution and study its impact on baryonic acoustic oscillations within the signal. These calculations are essential for comparing the observed trispectra of nonlinear cosmic density fields with theoretical expectations.
}

\begin{document}

\renewcommand{\thefootnote}{\fnsymbol{footnote}}
\footnotetext[2]{sazyzy@mpa-garching.mpg.de}
\maketitle

\flushbottom
\section{Introduction}
\label{sec:intro}
Introducing symmetries and testing their realization in nature has played an important role in the success of modern physics. Parity, which is a discrete symmetry corresponding to the reflection of the system on all spatial coordinates, is an example of such a symmetry that has transformed our understanding of physics. The discovery of violation of this symmetry in weak interactions \cite{Wu_1957, Garwin:1957hc} dictated the structure of the standard model of particle physics. However, it is an open question if parity is respected on large, cosmological scales \cite{Komatsu:2022nvu,Shiraishi_2016,Cahn:2021ltp}.

There have been recent hints of parity violation in the cosmic microwave background (CMB) in the form of cosmic birefringence \cite{Minami_2020, Diego_2022, Eskilt_2022,Diego-Palazuelos:2025dmh}. If confirmed, this signal provides evidence of parity-violating couplings that uniformly rotate the plane of linear polarization of CMB photons across the sky, and could shed light on the nature of dark matter and dark energy \cite{Carroll_1998, Nakatsuka:2022epj, Fedderke_2019}. 

There have also been hints of parity violation in the large-scale structure (LSS) of the universe, in the four-point statistics of galaxy clustering from the BOSS and DESI surveys, with reported detection significance ranging from $2$--$10\sigma$ \cite{Philcox_2022, Hou_2023, Slepian:2025kbb}. However, these results are sensitive to analysis choices \cite{Krolewski:2024paz, Slepian:2025kbb} and may be biased by systematics or covariance misestimation. If this signal is genuine, it could be evidence of new parity-violating physics from the early universe, e.g., cosmic inflation \cite{Starobinsky:1980te,Guth:1980zm,Sato:1981qmu,Linde:1981mu,Albrecht:1982wi},
imprinted as non-Gaussian and parity-odd correlations in the primordial scalar curvature perturbations. Several models of inflation proposed in the literature can generate such primordial parity-violating correlations \cite{Anber_2010, Barnaby_2011, Cabass_2023, Shiraishi_2016,Stefanyszyn:2023qov}.

Assuming parity violation was imprinted on the primordial curvature perturbations by some unknown primordial source during inflation, these are conserved outside the horizon \cite{Salopek:1990jq,Weinberg_2003,Lyth:2004gb} and imprint on the matter perturbations that enter the horizon during CMB and LSS eras. Multiple studies have tested cosmological parity violation using two- and three-point correlation functions of vector and tensor quantities \cite{CarrollFieldJackiw1990, LueWangKamionkowski1999, GluscevicKamionkowski2010, KamionkowskiSouradeep2011, Alexander2008, BartoloMatarresePelosoShiraishi2015, SetoTaruya2007, YunesOShaughnessyOwenAlexander2010, MasuiPenTurok2017, JenksChoiLagosYunes2023, BiagettiOrlando2020, YuMotlochPenYuWangMoYangJing2020, ShimPenYuOkumura2024, Ragavendra:2025svk, Kurita:2025hmp}. However, a direct probe of cosmological parity violation through the statistics of \textit{scalar} matter density perturbations is only possible through their trispectrum \cite{Shiraishi_2016} --- the Fourier transform of their four-point correlation function --- or higher-order statistics.

Analyses of the parity-odd galaxy four-point function so far have either focused on model-independent searches for a signal with nonlinear, mock-based covariance estimation \cite{Hou_2023, Krolewski:2024paz, Slepian:2025kbb}, or assumed a specific model for the primordial trispectrum but restricted the analysis to linear scales \cite{Philcox_2022}. We can obtain stronger constraints by extending model-dependent searches down to nonlinear scales, but then we need to know how the parity-odd trispectrum evolves under nonlinear gravitational clustering. Nonlinear gravitational evolution will not generate a parity-odd trispectrum. However,  the shape of the parity-odd trispectrum will be nontrivially distorted through gravitational clustering if it is present in the primordial density field.
Interpreting and robustly quantifying the detection of such a signal requires studying this nonlinear evolution.

The purpose of this work is to give an analytical expression for the nonlinear parity-violating matter trispectrum to 1-loop order (i.e. to third order in matter density fluctuations) within the effective field theory of LSS (EFTofLSS) framework \cite{Baumann_2012,Carrasco_2012}. Our work provides a connection between theoretical models of primordial parity-violating physics and LSS observations, enabling extensions of the search for new primordial parity-violating physics down to smaller scales.

The remainder of this paper is structured as follows. In Section \ref{sec:powerspectrum}, we review the standard Eulerian perturbation theory (SPT) and the matter power spectrum at 1-loop order. We also introduce our Feynman rules for a diagrammatic representation of the correlation functions of the matter density perturbations. In Section \ref{sec:trispectrum}, we calculate the parity-violating matter trispectrum at 1-loop order, investigate the infrared (IR) and ultraviolet (UV) limits, and find the UV counterterms. In Section \ref{sec:template}, we apply our trispectrum expression to a parity-violating primordial template and analyze our results. We illustrate the damping and shift of the baryonic acoustic oscillations (BAO) in the parity-violating part of the corresponding nonlinear matter trispectrum our parity-violating nonlinear matter trispectrum. We conclude in Section \ref{sec:conclusions}.

\section{Review of the nonlinear matter power spectrum}
\label{sec:powerspectrum}
\subsection{Standard Eulerian perturbation theory}
\label{sec:SPT}

In this section, we review the SPT of the matter field (also see \cite{Bernardeau_2002} for a comprehensive review). On scales larger than the baryonic Jeans scale, baryonic pressure can be neglected, and the baryons and cold dark matter can be treated as a single, pressureless fluid. We define the total matter density contrast, 
\begin{equation}
    \delta_m(\vec{x},\tau) = \frac{\rho_m(\vec{x},\tau) - \bar{\rho}_m(\tau)}{\bar{\rho}_m(\tau)} \, ,
\end{equation}
where $\rho_m(\vec{x},\tau)$ is the total matter density field, $\bar{\rho}_m(\tau)$ is its spatial average, $\vec{x}$ are the comoving coordinates, and $\tau$ is the conformal time.

On scales that are much smaller than the Hubble radius, relativistic effects of gravity can be neglected \cite{Peebles:1980yev}. In this regime, the matter evolves under Newtonian gravitational clustering on an expanding, cosmological background. This system is well-described by the continuity, Euler, and Poisson equations \cite{Peebles:1980yev}:
\begin{align}
    \label{eq:continuity}
    &\delta'_m+\vec{\nabla} \cdot [(1 + \delta_m)\vec{v}_m] = 0 \, , \\
    \label{eq:euler}
    &\vec{v}_m^{\, \prime} + (\vec{v}_m \cdot \vec{\nabla})\vec{v}_m = - \mathcal{H} \vec{v}_m - \vec{\nabla} \phi \, , \\
    \label{eq:poisson}
    &\nabla^2\phi = 4 \pi G a^2 \bar{\rho}_m\delta_m \, ,
\end{align}
where derivatives with respect to $\tau$ are denoted by primes. The conformal Hubble parameter is $\mathcal{H} = \ln(a)'$, where $a$ is the scale factor. The peculiar velocity field is $\vec{v}_m$. The Newtonian gravitational potential is denoted by $\phi$, and $\vec{\nabla}$ is the derivative with respect to $\vec{x}$. The linear solution to the fluid equations can be written as
\begin{equation}
    \delta_m^{(1)}(\vec{k}, \tau) = D(\tau) \delta_m^{(1)}(\vec{k}) \, ,
\end{equation}
where $D(\tau)$ is the growth factor given by $D(\tau) \propto \tau^2 $ during the matter-dominated era (or in terms of the scale factor, $D(a) \propto a$). 

Sources of parity violation in the early universe can imprint on the primordial curvature perturbations, which are conserved outside the horizon  \cite{Salopek:1990jq,Weinberg_2003,Lyth:2004gb}. For perturbations that re-enter the horizon during the matter-dominated era, the conserved curvature perturbations are related to the gravitational potential through $\phi(\vec{k}) = -(3/5) \zeta(\vec{k})$ \cite{Kodama:1984ziu}. These are then related to the matter density perturbations through the Poisson equation, shown in Eq.~\eqref{eq:poisson}. The wavenumber dependence of the linear matter density contrast is given by
\begin{align}
    \label{eq:linmatter}
    \delta_m^{(1)}(\vec{k}) =\mathcal{M} (k)\zeta(\vec{k})\, , \\
    \label{eq:transfer}
   \mathcal{M}(k) = \frac{2 k^2 \mathcal{T}(k)}{5 H_0^2 \Omega_m} \, ,
\end{align}
which comes, on the one hand, from the initial conditions (curvature perturbations, $\zeta(\vec{k})$), and on the other hand, from the transfer function, $\mathcal{T}(k)$, encoding scale-dependent effects during linear gravitational evolution. Here, $\Omega_m$ is the matter density parameter and $H_0$ the Hubble constant.

The nonlinear equations can be solved using SPT. In this framework, the matter density contrast in Fourier space can be expanded as 
\begin{equation}
    \label{eq:matter}
    \delta_m (\vec{k}, \tau) = \delta_m^{(1)}(\vec{k}, \tau) + \delta_m^{(2)}(\vec{k}, \tau) + \delta_m^{(3)}(\vec{k}, \tau) + \dots \, .
\end{equation}
At each order of perturbation theory, $\delta_m^{(n)}$ can be well approximated\footnote{The full system of equations is not exactly separable.} as $n$th powers of the linear matter density contrast, $\delta_m^{(1)}$
\cite{Goroff:1986ep}
\begin{equation}
    \label{eq:nonlinmatter}
    \begin{aligned}
    \delta_m^{(n)}(\vec{k}, \tau) &\simeq D^n(\tau) \delta_m^{(n)}(\vec{k})\\
    &=D^n(\tau)\int_{\vec{k}_1,\dots,\vec{k}_n} (2 \mathbf{\pi})^3 \delta_D(\vec{k}-\vec{k}_{12\dots n}) F_n^{(s)}(\vec{k}_1,\dots,\vec{k}_n) \delta_m^{(1)}(\vec{k}_1)\dots\delta_m^{(1)}(\vec{k}_n) \, ,
    \end{aligned}
\end{equation}
where we have used the notation,
\begin{equation}
\int_{\vec{q}} \equiv\int \frac{d^3q}{(2 \pi)^3}\,,\quad \vec{k}_{12 \dots n} \equiv \vec{k}_{1} + \vec{k}_{2} +\dots + \vec{k}_{n}\,,
\end{equation}
and $F_n^{(s)}$ are the SPT kernels that can be found recursively by plugging in the ansatz from Eq.~\eqref{eq:nonlinmatter} into the fluid equations, Eqs.~\eqref{eq:continuity}--\eqref{eq:poisson}. They are given in Appendix \ref{app:kernels} up to third order. 

\subsection{Matter power spectrum}
\label{sec:power}
Having found a solution for the matter density contrast in SPT, we can compute the nonlinear matter power spectrum, $P_m(k)$, defined as
\begin{equation}
    \begin{aligned}
        \langle \delta_m(\vec{k}, \tau)\delta_m(\vec{k}', \tau) \rangle
        &= (2\pi)^3\delta_D(\vec{k} + \vec{k}')P_m(k, \tau) \, .
    \end{aligned}
\end{equation}
In SPT, the power spectrum at 1-loop order is obtained by expanding the matter density contrasts up to third order using Eq.~\eqref{eq:matter}. For each order in perturbation theory, we define
\begin{equation}
    \begin{aligned}
        \langle \delta_m^{(a)}(\vec{k})\delta_m^{(b)}(\vec{k}') \rangle = (2\pi)^3\delta_D(\vec{k} + \vec{k}') P_{m}^{ab}(k) \, ,
    \end{aligned} 
\end{equation}   
where $a$ and $b$ are positive integers representing the order of perturbation theory. The nonlinear matter power spectrum up to 1-loop order can then be written as
\begin{equation}
    P_m(k, \tau) = D^2(\tau)P_m^{11}(k)+ 2 D^3(\tau)P_m^{12}(k) + D^4(\tau)\left[2 P_m^{13}(k)+ P_m^{22}(k)\right] +\mathcal{O}((\delta_m^{(1)})^5) \, ,
\end{equation}
where 
\begin{align}
    \label{eq:p11}
    P_m^{11}(k) &=\mathcal{M}^2(k) P_{\zeta}(k) \, , \\
    \label{eq:p12}
    P_m^{12}(k) &= \int_{\vec{q}} F_2^{(s)}(\vec{q}, \vec{k}-\vec{q}) B_m^{111}(\vec{q}, -\vec{k}, \vec{k}-
    \vec{q}) \, , \\
    \label{eq:p22}
    P_m^{22}(k) &= 2 \int_{\vec{q}} \left[ F_2^{(s)}(\vec{q}, \vec{k}-\vec{q}) \right]^2 P_m^{11}(q)P_m^{11}(|\vec{k}-\vec{q}|) \, , \\
    \label{eq:p13}
    P_m^{13}(k) &= 3P_m^{11}(k) \int_{\vec{q}} F_3^{(s)}(\vec{q}, -\vec{q}, \vec{k}) P_m^{11}(q) \, .
\end{align}
Here, $P_{\zeta}$ is the power spectrum of the primordial curvature perturbations and $B_m^{111}$ is the linear matter bispectrum coming from primordial non-Gaussianity (PNG) \cite{Bartolo:2004if}.

Using diagrammatic representations with Feynman-like rules, as defined in Figure~\ref{fig:Feynman}, is often useful for writing down each contributing term. Figure~\ref{fig:powerspectrum} shows diagrammatic representations of Eqs.~\eqref{eq:p11}--\eqref{eq:p13}.

\begin{figure}
    \centering
    \begin{subfigure}[b]{0.3\textwidth}
        \centering
        \begin{tikzpicture}
            \draw[dashed, thick] (-1,0) -- (1,0);
            \fill[white] (0,0) circle (0.1);
            \draw[pattern = crosshatch,thick] (0,0) circle (0.1);
        \end{tikzpicture}
        \newline
        \caption{$P_{\zeta}(k)$}
    \end{subfigure}
    \begin{subfigure}[b]{0.3\textwidth}
        \centering
        \begin{tikzpicture}
            \draw[solid, thick] (-1,0) -- (0,0);
            \draw[dashed, thick] (0,0) -- (1,0);
        \end{tikzpicture}
        \newline
        \caption{$\mathcal{M}(k)$}
    \end{subfigure}
    \begin{subfigure}[b]{0.3\textwidth}
        \centering
        \begin{tikzpicture}
            \draw[dashed, thick] (0,0) circle (0.5);
            \draw[solid, thick] (0.5,0) arc (0:90:0.5);
            \fill[white] (-0.35,-0.35) circle (0.1);
            \draw[pattern = crosshatch, thick] (-0.35,-0.35) circle (0.1);
        \end{tikzpicture}
        \caption{$\int_{\vec{q}}$}
    \end{subfigure}
    \newline
    \newline
    \begin{subfigure}[b]{0.3\textwidth}
        \centering
        \begin{tikzpicture}
            \draw[solid, thick] (-1,0) -- (-0.4,0);
            \draw[solid, thick] (-0.4,0) -- (0.4,0.4);
            \draw[solid, thick] (-0.4,0) -- (0.4,-0.4);
        \end{tikzpicture}
        \caption{$F_2^{(s)}(\vec{k}_1, \vec{k}_2)$}
    \end{subfigure}
    \begin{subfigure}[b]{0.3\textwidth}
        \centering
        \begin{tikzpicture}
            \draw[solid, thick] (-1,0) -- (0.4,0);
            \draw[solid, thick] (-0.4,0) -- (0.4,0.4);
            \draw[solid, thick] (-0.4,0) -- (0.4,-0.4);
        \end{tikzpicture}
        \caption{$F_3^{(s)}(\vec{k}_1, \vec{k}_2, \vec{k}_3)$}
    \end{subfigure}
    \caption{Feynman rules for diagrammatic representations of correlation functions. The dashed circle represents the full primordial power spectrum with general PNG, including loops.}
    \label{fig:Feynman}
\end{figure}
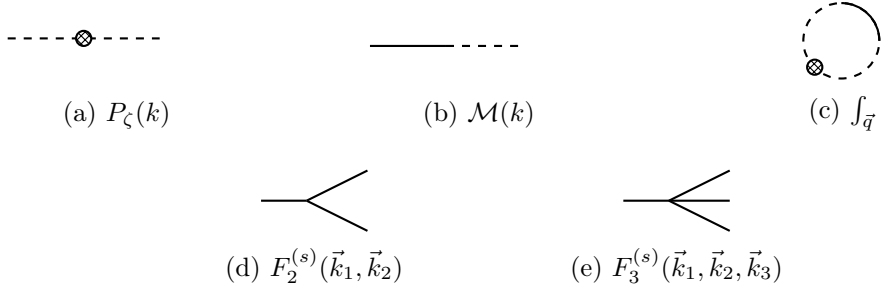
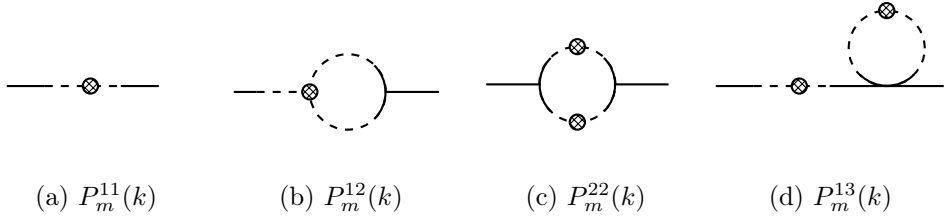
\begin{figure}
    \centering
    \begin{subfigure}[b]{0.2\textwidth}
        \centering
        \begin{tikzpicture}
            \draw[solid, thick] (-1,0) -- (-0.5,0);
            \draw[dashed, thick] (-0.5,0) -- (0.5,0);
            \draw[solid, thick] (0.5,0) -- (1,0);
            \fill[white] (0.1,0) circle (0.1);
            \draw[pattern = crosshatch, thick] (0.1,0) circle (0.1);
        \end{tikzpicture}
        \newline
        \newline
        \caption{$P_m^{11}(k)$}
    \end{subfigure}
    \begin{subfigure}[b]{0.2\textwidth}
        \centering
        \begin{tikzpicture}
            \draw[solid, thick] (-1.5,0) -- (-1.2,0);
            \draw[dashed, thick] (-1.2,0) -- (-0.5,0);
            \draw[dashed, thick] (0,0) circle (0.5);
            \draw[solid, thick] (0.35,0.35) arc (45:-45:0.5);
            \draw[solid, thick] (0.5,0) -- (1.2,0);
            \fill[white] (-0.5,0) circle (0.1);
            \draw[pattern = crosshatch, thick] (-0.5,0) circle (0.1);
        \end{tikzpicture}
        \newline
        \caption{$P_m^{12}(k)$}
    \end{subfigure}
    \begin{subfigure}[b]{0.2\textwidth}
        \centering
        \begin{tikzpicture}
            \draw[solid, thick] (-1.2,0) -- (-0.5,0);
            \draw[dashed, thick] (0,0) circle (0.5);
            \draw[solid, thick] (-0.35,0.35) arc (135:225:0.5);
            \draw[solid, thick] (0.35,0.35) arc (45:-45:0.5);
            \draw[solid, thick] (0.5,0) -- (1.2,0);
            \fill[white] (0, -0.5) circle (0.1);
            \fill[white] (0, 0.5) circle (0.1);
            \draw[pattern = crosshatch, thick] (0, -0.5) circle (0.1);
            \draw[pattern = crosshatch, thick] (0, 0.5) circle (0.1);
        \end{tikzpicture}
        \newline
        \caption{$P_m^{22}(k)$}
    \end{subfigure}
    \begin{subfigure}[b]{0.2\textwidth}
        \centering
        \begin{tikzpicture}
            \draw[solid, thick] (-1,0) -- (-0.5,0);
            \draw[dashed, thick] (-0.5,0) -- (0.5,0);
            \draw[solid, thick] (0.5,0) -- (2,0);
            \draw[solid, thick] (0.9, 0.15)  arc (225:315:0.5);
            \draw[dashed, thick] (1.25, 0.5) circle (0.5);
            \fill[white] (0.1,0) circle (0.1);
            \fill[white] (1.25, 1) circle (0.1);
            \draw[pattern = crosshatch, thick] (0.1,0) circle (0.1);
            \draw[pattern = crosshatch, thick] (1.25, 1) circle (0.1);
        \end{tikzpicture}
        \newline
        \newline
        \caption{$P_m^{13}(k)$}
    \end{subfigure}
    \caption{Diagrammatic representations of the matter power spectrum with PNG up to 1-loop order, corresponding to Eqs.~\eqref{eq:p11}--\eqref{eq:p13}.}
    \label{fig:powerspectrum}
\end{figure}
We now review the IR and UV limits of the 1-loop power spectrum. The IR modes ($q \ll k$) lead to a homogeneous shift of the system and do not contribute to the nonlinear corrections to the correlation functions due to the equivalence principle \cite{Jain:1995kx,Scoccimarro:1995if}. The IR limit of $P_m^{12}$ vanishes at leading order in $q/k$  
\begin{equation}
    \label{eq:P12_IR}
    \begin{aligned}
        P_m^{12}(k)\Big{|}_{IR} &\simeq \int \frac{d\Omega_{\hat{q}}}{4 \pi} \int_{q \ll k} \frac{dq \, q^2}{2\pi^2} \left(\frac{1}{2} \frac{k}{q} (\hat{q} \cdot \hat{k}) \right) B_m^{111}(\vec{q}, \vec{k}, -\vec{k}) \,  \\
        &= 0  \, ,
    \end{aligned}
\end{equation}
where $d\Omega_{\hat{q}}$ is the differential solid angle for angular integration over all $\hat{q}$ directions. This vanishes because $B_m^{111}(\vec{q}, \vec{k}, -\vec{k})$ is symmetric in $\vec{k} \rightarrow-\vec{k}$, whereas the leading contribution of the $F_2^{(s)}$ kernel is antisymmetric. 

For the other two terms, we have \cite{Vishniac_1983, Senatore_2015}
\begin{align}
    \label{eq:P22_IR}
    P_m^{22}(k)\Big{|}_{IR} &= \frac{2}{3}\frac{k^2 }{8 \pi^2} P_m^{11}(k) \int_{q \ll k} dq \, P_m^{11}(q) \, ,\\
    \label{eq:P13_IR}
    P_m^{13}(k)\Big{|}_{IR} &= -\frac{1}{3}\frac{k^2}{8 \pi^2} P_m^{11}(k) \int_{q \ll k} dq \, P_m^{11}(q) \, ,
\end{align}
which result in an IR cancellation
\begin{equation}
    2P_m^{13}(k)\Big{|}_{IR} + P_m^{22}(k)\Big{|}_{IR} = 0 \, .
\end{equation}
Therefore, we have confirmed that large-scale modes do not contribute to the nonlinear part of the correlation function. 

In the framework of EFTofLSS \cite{Baumann_2012,Carrasco_2012}, the loop integrals are truncated at a scale $q<\Lambda$ below the nonlinear scale $k_{nl}$, i.e., $\Lambda<k_{nl}$, where perturbation theory breaks down, and UV counterterms are added that contain the information about the short-scale physics at $q>\Lambda$. To find the form of the counterterms, we take the UV limits ($q\gg k$) of the 1-loop power spectrum and find \cite{Baumann_2012, Carrasco_2012, Baldauf_2015, ivanov_2022}
\begin{align}
        \label{eq:P12UV}
        P_{m}^{12}(k)\Big{|}_{UV} &\propto \int_{q\gg k} dq \ q^2\frac{k^2}{q^2} B_m^{111}(\vec{q}, \vec{k}, -\vec{q}) \, ,\\ 
        \label{eq:P13UV}
        P_m^{13}(k)\Big{|}_{UV}&\propto P_m^{11}(k) \int_{q\gg k} dq \ q^2\frac{k^2}{q^2} P_m^{11}(q) \propto k^2 P_m^{11}(k) \, ,\\
        \label{eq:P22UV}
        P_m^{22}(k)\Big{|}_{UV} &\propto \int_{q\gg k} dq \ q^2 \left( \frac{k^2}{q^2} \right)^2 \left[P_m^{11}(q)\right]^2 \propto k^4 \, ,
\end{align}
where we used the properties of the kernels given in Appendix \ref{app:kernels}. For $P^{12}_m(k)$, the UV scaling and hence the counterterm are dependent on the shape of PNG. For perturbative PNG, see Ref.~\cite{Assassi_2015}. The counterterm of $P_m^{22}(k)$ must scale as $k^4$, but this is, in general, small and can be neglected. Finally, the counterterm of $P_m^{13}(k)$ is 
\begin{equation}
    \label{eq:P13ct}
    P_m^{13}(k, \tau) \rightarrow P_m^{13}(k, \tau) \Big{|}_{\Lambda} - c_{s,\Lambda}^2(\tau)k^2 P_m^{11}(k) \, ,
\end{equation}
where the superscript ${\Lambda}$ corresponds to a truncation of the loop integral and $c_{s,\Lambda}$ is the effective sound speed coming from the effective stress tensor after smoothing the density fields \cite{Carrasco_2012}. As we will show below, the form of this counterterm is also relevant for the trispectrum.

\section{Parity-violating matter trispectrum}
\label{sec:trispectrum}
In this section, we derive the nonlinear parity-odd matter trispectrum up to 1-loop order. Since the matter power spectrum and bispectrum are insensitive to parity due to rotational invariance, cosmological parity violation can only be probed through the matter trispectrum (or higher-order statistics) \cite{Shiraishi_2016}. The matter trispectrum, $T_m$, is defined as the connected part of the Fourier transform of the four-point correlation function of the matter density contrast,
\begin{equation}
    \langle\delta_m(\vec{k}_1, \tau)\delta_m(\vec{k}_2, \tau)\delta_m(\vec{k}_3, \tau)\delta_m(\vec{k}_4, \tau)\rangle_c = (2\pi)^3\delta_D(\vec{k}_{1234})T_{m}(\vec{k}_1,\vec{k}_2, \vec{k}_3, \vec{k}_4, \tau) \, ,
\end{equation}
where, by $\langle\dots \rangle_c$, we refer to the connected part of the correlation function. The Dirac delta in the trispectrum enforces momentum conservation, $\vec{k}_{1234} =0$. Geometrically, this means that the four wave vectors form a closed shape in the three-dimensional Fourier space, corresponding to a tetrahedron. 

A parity transformation of the Fourier transform of real density fields corresponds to complex conjugation. Therefore, we have for the trispectrum
\begin{equation}
    \hat{P}: T_{m}(\vec{k}_1,\vec{k}_2, \vec{k}_3, \vec{k}_4, \tau) \rightarrow T_{m}(-\vec{k}_1,-\vec{k}_2, -\vec{k}_3, -\vec{k}_4, \tau) = T^*_{m}(\vec{k}_1,\vec{k}_2, \vec{k}_3, \vec{k}_4, \tau) \, , 
\end{equation}
where $\hat{P}$ represents a parity transformation. Only the imaginary part of the trispectrum is sensitive to parity. The trispectrum depends on the wave vectors through the geometrical properties of the tetrahedron they form, which can be in either a left-handed or right-handed configuration. The difference between the trispectrum for left- and right-handed configurations corresponds to the parity-odd trispectrum \cite{Jamieson:2024mau}. Hence, a non-zero parity-odd trispectrum implies a violation of parity symmetry in the scalar density contrast. Therefore, we decompose the trispectrum into the parity-even part, $T_{m,+}$, and the parity-odd part, $T_{m,-}$, in which we are interested in this work. So we can write
\begin{equation}
    T_{m}(\vec{k}_1,\vec{k}_2, \vec{k}_3, \vec{k}_4, \tau) =T_{m,+}(\vec{k}_1,\vec{k}_2, \vec{k}_3, \vec{k}_4, \tau) + i \, T_{m,-}(\vec{k}_1,\vec{k}_2, \vec{k}_3, \vec{k}_4,\tau) \, .
\end{equation}

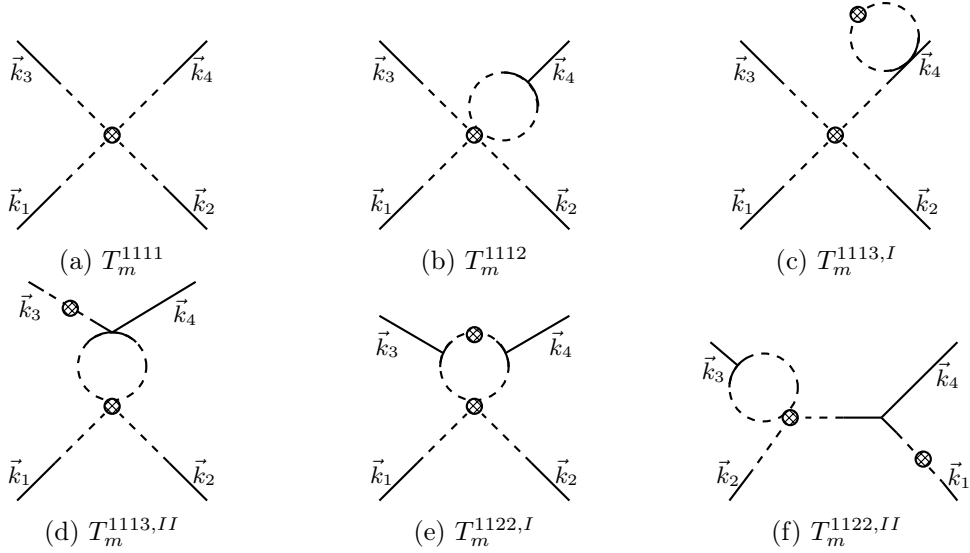
\begin{figure}
    \centering
    \begin{subfigure}[b]{0.3\textwidth} 
        \centering 
        \begin{tikzpicture} 
            \draw[solid, thick] (1.25,1.25) -- (0.75,0.75) node[very near start, below] {\footnotesize $\vec{k}_4$};
            \draw[dashed, thick] (0.75,0.75) -- (0.1,0.1);
            \draw[solid, thick] (-1.25,1.25) -- (-0.75,0.75) node[very near start, below] {\footnotesize $\vec{k}_3$};
            \draw[dashed, thick] (-0.75,0.75) -- (-0.1,0.1);
            \draw[solid, thick] (1.25,-1.25) -- (0.75,-0.75) node[very near start, above] {\footnotesize $\, \vec{k}_2$};
            \draw[dashed, thick] (0.75,-0.75) -- (0.1,-0.1);
            \draw[solid, thick] (-1.25,-1.25) -- (-0.75,-0.75)  node[very near start, above] {\footnotesize $\vec{k}_1 \, $};
            \draw[dashed, thick] (-0.75,-0.75) -- (-0.1,-0.1);
            \fill[white] (0,0) circle (0.1);
            \draw[pattern = crosshatch, thick] (0,0) circle (0.1);
        \end{tikzpicture} 
        \caption{$T_m^{1111}$} 
        \label{fig:T1111} 
    \end{subfigure}
    \begin{subfigure}[b]{0.3\textwidth}
        \centering 
        \begin{tikzpicture}
            \draw[solid, thick] (1.25,1.25) -- (0.7,0.7) node[very near start, below] {\footnotesize $\vec{k}_4$};
            \draw[solid, thick] (0.39,0.83) arc (90:0:0.45);
            \draw[solid, thick] (-1.25,1.25) -- (-0.75,0.75) node[very near start, below] {\footnotesize $\vec{k}_3$};
            \draw[dashed, thick] (-0.75,0.75) -- (-0.1,0.1);
            \draw[solid, thick] (1.25,-1.25) -- (0.75,-0.75) node[very near start, above] {\footnotesize $\, \vec{k}_2$};
            \draw[dashed, thick] (0.75,-0.75) -- (0.1,-0.1);
            \draw[solid, thick] (-1.25,-1.25) -- (-0.75,-0.75)  node[very near start, above] {\footnotesize $\vec{k}_1 \, $};
            \draw[dashed, thick] (-0.75,-0.75) -- (-0.1,-0.1);
            \draw[dashed, thick] (0.38,0.38) circle (0.45);
            \fill[white] (0,0) circle (0.1);
            \draw[pattern = crosshatch, thick] (0,0) circle (0.1);
        \end{tikzpicture}
        \caption{$T_m^{1112}$} 
        \label{fig:T1112} 
    \end{subfigure}
    \begin{subfigure}[b]{0.3\textwidth}
        \centering 
        \begin{tikzpicture}
            \draw[solid, thick] (1.25,1.25) -- (0.75,0.75) node[pos=0, below] {\footnotesize $\vec{k}_4$};
            \draw[dashed, thick] (0.75,0.75) -- (0.1,0.1);
            \draw[solid, thick] (-1.25,1.25) -- (-0.75,0.75) node[very near start, below] {\footnotesize $\vec{k}_3$};
            \draw[dashed, thick] (-0.75,0.75) -- (-0.1,0.1);
            \draw[solid, thick] (1.25,-1.25) -- (0.75,-0.75) node[very near start, above] {\footnotesize $\, \vec{k}_2$};
            \draw[dashed, thick] (0.75,-0.75) -- (0.1,-0.1);
            \draw[solid, thick] (-1.25,-1.25) -- (-0.75,-0.75)  node[very near start, above] {\footnotesize $\vec{k}_1 \, $};
            \draw[dashed, thick] (-0.75,-0.75) -- (-0.1,-0.1);
            \draw[dashed, thick] (0.65, 1.3) circle (0.45);
            \draw[solid, thick] (0.72, 0.85) arc (280:360:0.45);
            \fill[white] (0.3,1.6) circle (0.1);
            \draw[pattern = crosshatch, thick] (0.3,1.6) circle (0.1);
            \fill[white] (0,0) circle (0.1);
            \draw[pattern = crosshatch, thick] (0,0) circle (0.1);
        \end{tikzpicture}
        \caption{$T_m^{1113,I}$} 
        \label{fig:T1113I} 
    \end{subfigure}
    \begin{subfigure}[b]{0.3\textwidth}
        \centering 
        \begin{tikzpicture}
            \draw[dashed, thick] (0,0.53) circle (0.45);
            \draw[solid, thick] (-0.32,0.85) arc (135:45:0.45);
            \draw[solid, thick] (0, 0.98) -- (1.1,1.65) node[very near end, below] {\footnotesize $\vec{k}_4$};
            \draw[solid, thick] (0, 0.98) -- (-0.2,1.1);
            \draw[dashed, thick] (-0.2, 1.1) -- (-0.9, 1.53);
            \draw[solid, thick] (-0.9, 1.53) -- (-1.1,1.65) node[very near end, below] {\footnotesize $\vec{k}_3$};
            \draw[solid, thick] (1.25,-1.25) -- (0.75,-0.75) node[very near start, above] {\footnotesize $\, \vec{k}_2$};
            \draw[dashed, thick] (0.75,-0.75) -- (0.1,-0.1);
            \draw[solid, thick] (-1.25,-1.25) -- (-0.75,-0.75) node[very near start, above] {\footnotesize $\vec{k}_1 \, $};
            \draw[dashed, thick] (-0.75,-0.75) -- (-0.1,-0.1);
            \fill[white] (-0.55,1.3) circle (0.1);
            \draw[pattern = crosshatch, thick] (-0.55,1.3) circle (0.1);
            \fill[white] (0,0) circle (0.1);
            \draw[pattern = crosshatch, thick] (0,0) circle (0.1);
        \end{tikzpicture}
        \caption{$T_m^{1113,II}$} 
        \label{fig:T1113II} 
    \end{subfigure}
    \begin{subfigure}[b]{0.3\textwidth}
        \centering 
        \begin{tikzpicture}
            \draw[dashed, thick] (0,0.53) circle (0.45);
            \draw[solid, thick] (-0.45,0.52) arc (180:130:0.45);
            \draw[solid, thick] (0.45,0.52) arc (0:50:0.45);
            \draw[solid, thick] (0.4, 0.7) -- (1.25,1.2) node[very near end, below] {\footnotesize $\vec{k}_4$};
            \draw[solid, thick] (-0.4, 0.7) -- (-1.25,1.2) node[very near end, below] {\footnotesize $\vec{k}_3$};
            \draw[solid, thick] (1.25,-1.25) -- (0.75,-0.75) node[very near start, above] {\footnotesize $\, \vec{k}_2$};
            \draw[dashed, thick] (0.75,-0.75) -- (0.1,-0.1);
            \draw[solid, thick] (-1.25,-1.25) -- (-0.75,-0.75) node[very near start, above] {\footnotesize $\vec{k}_1 \, $};
            \draw[dashed, thick] (-0.75,-0.75) -- (-0.1,-0.1);
            \fill[white] (0,0.95) circle (0.1);
            \draw[pattern = crosshatch, thick] (0,0.95) circle (0.1);
            \fill[white] (0,0) circle (0.1);
            \draw[pattern = crosshatch, thick] (0,0) circle (0.1);
        \end{tikzpicture}
        \caption{$T_m^{1122,I}$} 
        \label{fig:T1122I} 
    \end{subfigure}
    \begin{subfigure}[b]{0.3\textwidth}
        \centering 
        \begin{tikzpicture}
            \draw[dashed, thick] (-1.05,0.4) circle (0.45);
            \draw[solid, thick] (-1.5,0.4) arc (180:120:0.45);
            \draw[solid, thick] (-1.4, 0.7) -- (-1.75,1.0) node[very near end, below] {\footnotesize $\vec{k}_3$};
            \draw[dashed, thick] (0,0) -- (-0.5, 0);
            \draw[solid, thick] (0,0) -- (0.5,0);
            \draw[solid, thick] (0.5,0) -- (1.5,1) node[very near end, below] {\footnotesize $\vec{k}_4$};
            \draw[solid, thick] (0.5, 0) -- (0.7, -0.2);
            \draw[dashed, thick] (0.7, -0.2) -- (1.3, -0.85);
            \draw[solid, thick] (1.3,-0.85) -- (1.5, -1.1) node[very near end, above] {\footnotesize $\ \vec{k}_1$};
            \draw[dashed, thick] (-0.7,0) -- (-1.2,-0.7) ;
            \draw[solid, thick] (-1.2,-0.7) -- (-1.5,-1.1) node[very near end, above] {\footnotesize $\vec{k}_2 \ $};
            \fill[white] (-0.7,0) circle (0.1);
            \draw[pattern = crosshatch, thick] (-0.7,0) circle (0.1);
            \fill[white] (1.05,-0.55) circle (0.1);
            \draw[pattern = crosshatch, thick] (1.05,-0.55) circle (0.1);
        \end{tikzpicture}
        \caption{$T_m^{1122,II}$} 
        \label{fig:T1122II} 
    \end{subfigure}
    \caption{Diagrammatic representations of the trispectrum from Eq.~\eqref{eq:trispectrum} with PNG represented by the dashed circles. Diagram (f) does not contribute to the parity-odd trispectrum since it contains the parity-odd bispectrum.}
    \label{fig:trispectrum}
\end{figure}

For convenience, we ignore the time dependence in this section, since it only corresponds to a multiplication by powers of the growth factor. In order to calculate $T_{m,-}$, similar to Section~\ref{sec:power}, we expand the matter density fields in perturbation theory. For each nonlinear trispectrum contribution, we define
\begin{equation}
    \langle\delta_m^{(a)}(\vec{k}_1)\delta_m^{(b)}(\vec{k}_2)\delta_m^{(c)}(\vec{k}_3)\delta_m^{(d)}(\vec{k}_4) \rangle_c = (2\pi)^3 \delta_D(\vec{k}_{1234}) T_{m}^{abcd}(\vec{k}_1, \vec{k}_2, \vec{k}_3, \vec{k}_4) \, ,
\end{equation}
where $a$, $b$, $c$, and $d$ are positive integers corresponding to the order of perturbation theory. Each of the $T_m^{abcd}$ can be split into the parity-even and parity-odd parts, $T_m^{abcd} = T_{m,+}^{abcd} + i T_{m,-}^{abcd}$. The parity-odd trispectrum can then be expanded in SPT as
\begin{equation}
    \label{eq:trispectrum}
    \begin{aligned}
        T_{m,-}(\vec{k}_1, \vec{k}_2, \vec{k}_3, \vec{k}_4, \tau) =& D^4(\tau) \, T_{m,-}^{1111}(\vec{k}_1, \vec{k}_2, \vec{k}_3, \vec{k}_4) \\
        &+ D^5(\tau) \, (T_{m,-}^{1112}(\vec{k}_1, \vec{k}_2, \vec{k}_3, \vec{k}_4) \ + 3 \ \rm{perm.}) \\
        &+ D^6(\tau) \, (T_{m,-}^{1113}(\vec{k}_1, \vec{k}_2, \vec{k}_3, \vec{k}_4) \ + 3 \ \rm{perm.}) \\
        &+ D^6(\tau) \, (T_{m,-}^{1122}(\vec{k}_1, \vec{k}_2, \vec{k}_3, \vec{k}_4) \ + 5 \ \rm{perm.}) \\
        &+ \mathcal{O}((\delta_m^{(1)})^7) \, ,
    \end{aligned}
\end{equation}
where ``$\rm{perm.}$'' corresponds to permutations of $(a, b,c,d)$. Figure~\ref{fig:trispectrum} shows the diagrammatic representations of each term. 

The leading-order parity-odd matter trispectrum is related to the parity-odd trispectrum of primordial curvature perturbations, $T_{\zeta,-}$, by the transfer functions and linear growth factors (omitted here),
\begin{equation}
    \label{eq:Tlin}
    T_{m,-}^{1111}(\vec{k}_1, \vec{k}_2, \vec{k}_3, \vec{k}_4) \color{black} =\mathcal{M}(k_1)\mathcal{M}(k_2)\mathcal{M}(k_3) \mathcal{M}(k_4) \ T_{\zeta,-}(\vec{k}_1, \vec{k}_2, \vec{k}_3, \vec{k}_4) \, .
\end{equation}

In the remainder of this section, we will derive the nonlinear corrections to the trispectrum from Eq.~\eqref{eq:trispectrum}. What simplifies the derivation is the fact that the primordial parity-odd power spectrum and bispectrum are zero due to rotational invariance. So only the terms that contain higher-order primordial statistics contribute. The second term in Eq.~\eqref{eq:trispectrum} corresponding to diagram (b) in Figure~\ref{fig:trispectrum} is given by
\begin{equation}
    \begin{aligned}
        T_{m,-}^{1112}(&\vec{k}_1, \vec{k}_2, \vec{k}_3, \vec{k}_4) = \int_{\vec{q}} F_2^{(s)}(\vec{q}, \vec{k}_4 -\vec{q}) \ Q^{11111}_{m,-}(\vec{k}_1, \vec{k}_2, \vec{k}_3, \vec{q}, \vec{k}_4 - \vec{q}) \, ,
    \end{aligned}
\end{equation}
where we have defined $Q_{m,-}^{11111}$ as the Fourier transform of the linear parity-odd five-point function. This term is an analog of $P_m^{12}$ in Eq.~\eqref{eq:p12}. In Appendix~\ref{app:5pcf}, we give an example of a possible primordial template that produces a parity-odd five-point function. 

Next, we compute the third term in Eq.~\eqref{eq:trispectrum} corresponding to diagrams (c) and (d) in Figure~\ref{fig:trispectrum}.
For the nonlinear terms of order $\mathcal{O}((\delta_m^{(1)})^6)$, only the Wick contractions that contain a linear trispectrum and power spectrum contribute (another possibility would be a linear six-point function, but this would be of 2-loop order). We obtain
\begin{equation}
    \begin{aligned}
        \langle &\delta_m^{(1)}(\vec{k}_1)\delta_m^{(1)}(\vec{k}_2)\delta_m^{(1)}(\vec{k}_3)\delta_m^{(3)}(\vec{k}_4) \rangle_c \\
        &= \int_{\vec{q}, \vec{q}{\,}'} F_3^{(s)}(\vec{q}, \vec{q}{\,}', \vec{k}_4-\vec{q}-\vec{q}{\,}') \langle \delta_m^{(1)}(\vec{k}_1)\delta_m^{(1)}(\vec{k}_2)\delta_m^{(1)}(\vec{k}_3)\delta_m^{(1)}(\vec{q})\delta_m^{(1)}(\vec{q}{\,}') \delta_m^{(1)}(\vec{k}_4-\vec{q}- \vec{q}{\,}') \rangle_c \, .
    \end{aligned}
\end{equation}
There are two possible ways of Wick contraction yielding two terms that contribute to $T_{m,-}^{1113}$. We write
\begin{equation}
    T_{m,-}^{1113}(\vec{k}_1, \vec{k}_2, \vec{k}_3, \vec{k}_4) \equiv T_{m,-}^{1113,I}(\vec{k}_1, \vec{k}_2, \vec{k}_3, \vec{k}_4) + T_{m,-}^{1113, II}(\vec{k}_1, \vec{k}_2, \vec{k}_3, \vec{k}_4) \, .
\end{equation} 
The first term, $T_{m,-}^{1113,I}$, is obtained by contracting
$\delta_m^{(1)}(\vec{q})$ with $\delta_m^{(1)}(\vec{q}{\,}')$:
\begin{equation}
    \label{eq:T1113I}
    \begin{aligned}
        T_{m,-}^{1113,I}(\vec{k}_1, \vec{k}_2, \vec{k}_3, \vec{k}_4) &=  T_{m,-}^{1111}(\vec{k}_1, \vec{k}_2, \vec{k}_3, \vec{k}_4) \ 3 \int_{\vec{q}} F_3^{(s)}(\vec{q}, -\vec{q}, \vec{k}_4) P_m^{11}(q) \\ 
        &= T_{m,-}^{1111}(\vec{k}_1, \vec{k}_2, \vec{k}_3, \vec{k}_4) \frac{P_m^{13}(k_4)}{P_m^{11}(k_4)} \, .
    \end{aligned} 
\end{equation}
The second term, $T_{m,-}^{1113,II}$, is obtained by contracting
$\delta_m^{(1)}(\vec{k}_1)$, $\delta_m^{(1)}(\vec{k}_2)$ or $\delta_m^{(1)}(\vec{k}_3)$ with one of the three internal momenta (which gives a factor of 3):
\begin{equation}
    \label{eq:T1113II}
    \begin{aligned}
        &T_{m,-}^{1113,II}(\vec{k}_1, \vec{k}_2, \vec{k}_3, \vec{k}_4) \\
        &= 3 P_m^{11}(k_1)\int_{\vec{q}} F_3^{(s)}(\vec{q}, -\vec{k}_1, \vec{k}_1+\vec{k}_4-\vec{q}) \ T_{m,-}^{1111}(\vec{k}_2, \vec{k}_3, \vec{q}, \vec{k}_1+\vec{k}_4-\vec{q})
        + (\vec{k}_1 \leftrightarrow \vec{k}_2) + (\vec{k}_1 \leftrightarrow \vec{k}_3) \, .
    \end{aligned}
\end{equation}

Finally, we compute the fourth term in Eq.~\eqref{eq:trispectrum} corresponding to diagrams (e) and (f) in Figure~\ref{fig:trispectrum}.
We proceed as before and write
\begin{equation}
    \begin{aligned}
        &\langle \delta_m^{(1)}(\vec{k}_1)\delta_m^{(1)}(\vec{k}_2)\delta_m^{(2)}(\vec{k}_3)\delta_m^{(2)}(\vec{k}_4) \rangle_c \\
        &=\int_{\vec{q}, \vec{q}{\,}'} F_2^{(s)}(\vec{q}, \vec{k}_3-\vec{q}) F_2^{(s)}(\vec{q}, \vec{k}_4-\vec{q}) \langle \delta_m^{(1)}(\vec{k}_1)\delta_m^{(1)}(\vec{k}_2)\delta_m^{(1)}(\vec{q})\delta_m^{(1)}(\vec{k}_3-\vec{q})\delta_m^{(1)}(\vec{q}{\,}')\delta_m^{(1)}(\vec{k}_4-\vec{q}{\,}') \rangle_c \, .
    \end{aligned}
\end{equation}
Again, we contract two internal momenta like $\delta_m^{(1)}(\vec{q})$ with $\delta_m^{(1)}(\vec{q}{\,}')$ (there are four possibilities, hence the factor of 4 below) and obtain
\begin{equation}
    \label{eq:T1122I}
    \begin{aligned}
        T_{m,-}^{1122,I}(&\vec{k}_1, \vec{k}_2, \vec{k}_3, \vec{k}_4) = 4 \int_{\vec{q}} F_2^{(s)}(\vec{q}, \vec{k}_3 -\vec{q}) F_2^{(s)}(-\vec{q}, \vec{k}_4 +\vec{q}) P_m^{11}(q) T_{m,-}^{1111}(\vec{k}_1, \vec{k}_2, \vec{k}_3 - \vec{q}, \vec{k}_4 + \vec{q}) \, .
    \end{aligned}
\end{equation}
We then contract one of the internal momenta either with $\delta_m^{(1)}(\vec{k}_1)$ or $\delta_m^{(1)}(\vec{k}_2)$ and obtain
\begin{equation}
    \begin{aligned}
        T_{m,-}^{1122,II}(\vec{k}_1, \vec{k}_2, \vec{k}_3, \vec{k}_4) =& \ 
        2 P_m^{11}(k_1) F_2^{(s)}(-\vec{k}_1, \vec{k}_1 + \vec{k}_3) \int_{\vec{q}} \Big(F_2^{(s)}(\vec{q}, \vec{k}_4 - \vec{q}) \\
        & \quad T_{m,-}^{1111}(\vec{k}_1 + \vec{k}_3, \vec{k}_2, \vec{q}, \vec{k}_4- \vec{q}) \Big)
        + (\vec{k}_1 \leftrightarrow \vec{k}_2) + (\vec{k}_3 \leftrightarrow \vec{k}_4) \\
        =& \ 2 P_m^{11}(k_1) F_2^{(s)}(-\vec{k}_1, \vec{k}_1 + \vec{k}_3) B_{m, -}^{112} (\vec{k}_1 + \vec{k}_3, \vec{k}_2, \vec{k}_4) \\
        & \quad + (\vec{k}_1 \leftrightarrow \vec{k}_2) + (\vec{k}_3 \leftrightarrow \vec{k}_4) \\
        =& \ 0 \, .
    \end{aligned}
\end{equation}
We find that the above term is proportional to the parity-odd bispectrum, $B_{m,-}^{112}$. However, since the parity-odd bispectrum is zero due to rotational invariance, this term also vanishes. 

Each nonlinear correction term is proportional to the linear parity-odd trispectrum from Eq.~\eqref{eq:Tlin}, which is directly related to the primordial parity-odd trispectrum, $T_{\zeta,-}$. Therefore, provided that there is a primordial source of parity violation, the nonlinear evolution can be directly computed by plugging in any parity-odd trispectrum of curvature perturbations, $T_{\zeta,-}$, in the above formulae. 

In the next two sections, we further investigate the integrals of the nonlinear trispectrum and examine how they behave at both short and large scales.

\subsection{IR limit}
As we saw in the case of the power spectrum in Section \ref{sec:power}, the large-scale modes ($q\ll k$) do not contribute to the 1-loop corrections of the correlation functions. In this section, we check this for the case of the parity-odd trispectrum defined in Eq.~\eqref{eq:trispectrum}. See Appendix \ref{app:IR} for more details and derivations. 

For $T_{m,-}^{1112}$, we find similarly to $P_m^{12}$ given in Eq.~\eqref{eq:P12_IR}
\begin{equation}
    \begin{aligned}
        T_{m,-}^{1112}(\vec{k}_1, \vec{k}_2, &\vec{k}_3, \vec{k}_4) \Big{|}_{IR} \\
        &\simeq \int \frac{d\Omega_{\hat{q}}}{4 \pi} \int_{q \ll k} \frac{dq \, q^2}{2\pi^2} \left(\frac{1}{2} \frac{1}{q} (\hat{q} \cdot \vec{k}_{4}) \right) Q_m^{11111}(\vec{k}_1, \vec{k}_2, \vec{k}_3, \vec{k}_4, \vec{q}) \, .
    \end{aligned}    
\end{equation}
Summing over all the permutations gives
\begin{equation}
    \begin{aligned}
        (T_{m,-}^{1112}(\vec{k}_1, \vec{k}_2, &\vec{k}_3, \vec{k}_4) + 3 \ \rm{perm.})\Big{|}_{IR} \\
        &\simeq \int \frac{d\Omega_{\hat{q}}}{4 \pi} \int_{q \ll k} \frac{dq \, q^2}{2\pi^2} \left(\frac{1}{2} \frac{1}{q} (\hat{q} \cdot \vec{k}_{1234}) \right) Q_m^{11111}(\vec{k}_1, \vec{k}_2, \vec{k}_3, \vec{k}_4, \vec{q}) \\
        &= 0 \, ,
    \end{aligned}    
\end{equation}
where, as before, ``perm.'' refers to which wave vector corresponds to the second-order perturbation. This term vanishes due to translational invariance. 

Next, we take the IR limit of $T_{m,-}^{1113,I}$. This is simple, since we already know the IR limit of $P_m^{13}$ from Eq.~\eqref{eq:P13_IR}. We have 
\begin{equation}
    \begin{aligned}
        T_{m,-}^{1113,I}(\vec{k}_1, \vec{k}_2, \vec{k}_3, \vec{k}_4) \Big{|}_{IR} =-\frac{1}{3} \frac{1}{(2 \pi)^2} k_4^2 \, T_{m,-}^{1111}(\vec{k}_1, \vec{k}_2, \vec{k}_3, \vec{k}_4) \int_{q \ll k} dq \, P_m^{11}(q) \, .
    \end{aligned}
\end{equation}
For all the permutations, we get
\begin{equation}
    \label{eq:1113IR}
    \begin{aligned}
        (T_{m,-}^{1113,I}(\vec{k}_1, \vec{k}_2, &\vec{k}_3, \vec{k}_4) + 3 \ \rm{perm.})\Big{|}_{IR} \\
        &=-\frac{1}{3} \frac{1}{(2 \pi)^2} (k_1^2 + k_2^2 + k_3^2 + k_4^2)T_{m,-}^{1111}(\vec{k}_1, \vec{k}_2, \vec{k}_3, \vec{k}_4) \int_{q \ll k} dq \, P_m^{11}(q) \, .
    \end{aligned}
\end{equation}
And for $T_{m,-}^{1122,I}$, we find 
\begin{equation}
    T_{m,-}^{1122,I}(\vec{k}_1, \vec{k}_2, \vec{k}_3, \vec{k}_4)\Big{|}_{IR} = -\frac{1}{3} \frac{1}{(2 \pi)^2}  2 (\vec{k}_3 \cdot \vec{k}_4) T_{m,-}^{1111}(\vec{k}_1, \vec{k}_2, \vec{k}_3, \vec{k}_4)  \int_{q \ll k} dq \, P_m^{11}(q) \, .
\end{equation}
Taking all the permutations gives
\begin{equation}
    \label{eq:1122IR}
    \begin{aligned}
        (T_{m,-}^{1122,I}(\vec{k}_1, &\vec{k}_2, \vec{k}_3, \vec{k}_4) + 5 \ \rm{perm.})\Big{|}_{IR} \\
        &=\frac{1}{3} \frac{1}{(2 \pi)^2} (k_1^2 + k_2^2 + k_3^2 + k_4^2)T_{m,-}^{1111}(\vec{k}_1, \vec{k}_2, \vec{k}_3, \vec{k}_4) \int_{q \ll k} dq \, P_m^{11}(q) \, .
    \end{aligned}
\end{equation}
Summing Eqs.~\eqref{eq:1113IR} and \eqref{eq:1122IR} results in an IR cancellation. 

For the remaining non-zero term, $T_{m,-}^{1113,II}$, we find
\begin{equation}
    \label{eq:1113II_IR}
    (T_{m,-}^{1113,II}(\vec{k}_1, \vec{k}_2, \vec{k}_3, \vec{k}_4) + 3 \ \rm{perm.})\Big{|}_{IR} = 0 \, .
\end{equation}
Hence, we find that the 1-loop contributions to the parity-odd trispectrum in the IR limit vanish, $T_{m,-}|_{IR} \rightarrow T^{1111}_{m,-}$. In conclusion, we have, for the first time, confirmed that the IR modes decouple from local physics in the parity-odd 1-loop trispectrum, as required by the equivalence principle.

\subsection{EFTofLSS counterterms}
Similarly to Section \ref{sec:power}, we find the form of the UV counterterms of the parity-odd trispectrum by investigating how the loop integrals behave at UV ($q \gg k$). Taking the UV limit in $T_{m,-}^{1112}$ gives
\begin{equation}
    T_{m,-}^{1112}(\vec{k}_1, \vec{k}_2, \vec{k}_3, \vec{k}_4) \Big{|}_{UV} \propto \int d \Omega_{\hat{q}} \int_{q\gg k_i} dq \ q^2 \frac{k_4^2}{q^2} Q_{m,-}^{1111}(\vec{k}_1, \vec{k}_2, \vec{k}_3, \vec{q}, - \vec{q}) \, ,
\end{equation}
where $k_i$ denote $k_1, k_2, k_3$ and $k_4$. Similar to the case of $P^{12}_m(k)$ in Eq.~\eqref{eq:P12UV}, the UV limit of this term and hence the shape of the counterterm depends on the primordial five-point function that is being considered. 

For $T_{m,-}^{1122,I}$ and $T_{m,-}^{1113,II}$, we get 
\begin{align}
    \label{eq:uv1122}
    &T_{m,-}^{1122,I}(\vec{k}_1, \vec{k}_2, \vec{k}_3, \vec{k}_4)\Big{|}_{UV} \propto \int d \Omega_{\hat{q}} \int_{q\gg k_i} dq \ q^2 \frac{k_3^2}{q^2} \frac{k_4^2}{q^2} P_m^{11}(q) T_{m,-}^{1111}(\vec{k}_1, \vec{k}_2, -\vec{q}, \vec{q})  \, ,\\
    \label{eq:uv1113}
    &T_{m,-}^{1113,II}(\vec{k}_1, \vec{k}_2, \vec{k}_3, \vec{k}_4)\Big{|}_{UV} \propto \int d \Omega_{\hat{q}} \int_{q\gg  k_i} dq \ q^2 \frac{k_4^2}{q^2} T_{m,-}^{1111}(\vec{k}_2, \vec{k}_3, \vec{q}, -\vec{q}) \, .
\end{align}
Since the only parity-odd quantity we can build from three vectors $(\vec{v}_i, \vec{v}_j, \vec{v}_k)$ is the triple product, $\vec{v}_i \cdot (\vec{v}_j \times \vec{v}_k)$, we must have $T_{m,-}^{1111}(\vec{k}_1, \vec{k}_2, -\vec{q}, \vec{q}) \propto \vec{q} \cdot (\vec{k}_1 \times \vec{k}_2)$. Due to isotropy, the integral over $\vec{q}$ can only depend on $\vec{k}_1$ and $\vec{k}_2$, and so the integral must be a linear combination of these two vectors. The triple products then vanish and as a result, both of the UV limits in Eqs.~\eqref{eq:uv1122} and \eqref{eq:uv1113} are zero. 

Finally, the only UV contribution comes from $T_{m,-}^{1113,I}$
\begin{equation}
    T_{m,-}^{1113,I}(\vec{k}_1, \vec{k}_2, \vec{k}_3, \vec{k}_4)\Big{|}_{UV} = T_{m,-}^{1111}(\vec{k}_1, \vec{k}_2, \vec{k}_3, \vec{k}_4) \frac{P_m^{13}(k_4)\Big{|}_{UV}}{P_m^{11}(k_4)} \, .
\end{equation}
Using Eq.~\eqref{eq:P13UV}, this can be written as
\begin{equation}
    (T_{m,-}^{1113,I}(\vec{k}_1, \vec{k}_2, \vec{k}_3, \vec{k}_4) + 3 \ \mathrm{perm.})\Big{|}_{UV} \propto  T_{m,-}^{1111}(\vec{k}_1, \vec{k}_2, \vec{k}_3, \vec{k}_4) \ (k_1^2+k_2^2+k_3^2+k_4^2) \int_{q\gg k_i} dq \ q^2 \frac{P_m^{11}(q)}{q^2} \, .
\end{equation}
This means that the counterterm must scale as $k^2$ and has a similar form to that in Eq.~\eqref{eq:P13ct}. The 1-loop EFTofLSS parity-odd trispectrum is then given by
\begin{equation}
    \label{eq:Teft}
    \begin{aligned}
        T^{EFT}_{m,-}(\vec{k}_1, \vec{k}_2, \vec{k}_3, \vec{k}_4, \tau) =& \ D^4(\tau) \, T_{m,-}^{1111}(\vec{k}_1, \vec{k}_2, \vec{k}_3, \vec{k}_4) \\ 
        &+ \ D^5(\tau) \, (T_{m,-}^{1112}(\vec{k}_1, \vec{k}_2, \vec{k}_3, \vec{k}_4)^{\Lambda} \ + 3 \ \rm{perm.}) \\
        &+ \ D^6(\tau) \, (T_{m,-}^{1113}(\vec{k}_1, \vec{k}_2, \vec{k}_3, \vec{k}_4)^{\Lambda} \ + 3 \ \rm{perm.}) \\
        &+ \ D^6(\tau) \, (T_{m,-}^{1122}(\vec{k}_1, \vec{k}_2, \vec{k}_3, \vec{k}_4)^{\Lambda} \ + 5 \ \rm{perm.}) \\ 
        &- \ D^4(\tau) \, c_{s,\Lambda}^2(\tau) \ (k_1^2+k_2^2+k_3^2+k_4^2) \ T_{m,-}^{1111}(\vec{k}_1, \vec{k}_2, \vec{k}_3, \vec{k}_4) \\
        &+ \ \mathcal{O}((\delta_m^{(1)})^7) \, .
    \end{aligned}
\end{equation}
This is the main result of this paper. Here, the superscript $\Lambda$ indicates a truncated loop integral at the scale $\Lambda < k_{nl}$, and $k_{nl}$ is the nonlinear scale where perturbation theory breaks down. We omit the PNG-dependent counterterm for $T_{m,-}^{1112}$ in the above formula.
\section{Application to primordial template}
\label{sec:template}
Eq.~\eqref{eq:Teft} is valid for any parity-odd primordial trispectrum. To illustrate the effect of nonlinear evolution, we apply our results to an example for parity-violating PNG given by \cite{coulton_2023}
\begin{equation}
    \label{eq:PNG}
    \zeta (\vec{x}) = \zeta_G(\vec{x}) + g_- \vec{\nabla} \zeta_G^{[\alpha]} (\vec{x})\cdot \left[\vec{\nabla} \zeta_G^{[\beta]}(\vec{x}) \times \nabla \zeta_G^{[\gamma]} (\vec{x})\right] \, .
\end{equation}
In Fourier space, this becomes
\begin{equation}
    \label{eq:phik}
    \begin{aligned}
        \zeta (\vec{k}) &= \zeta_G(\vec{k}) - i g_{-} \int_{\vec{q}_1, \vec{q}_2, \vec{q}_3} (2 \mathbf{\pi})^3 \delta^{(3)}_D(\vec{k}-\vec{q}_{123}) \zeta_G^{[\alpha]}(\vec{q_1}) \zeta_G^{[\beta]} (\vec{q}_2) \zeta_G^{[\gamma]} (\vec{q}_3) \vec{q}_1 \cdot (\vec{q}_2 \times \vec{q}_3)  \\
        &\equiv \zeta^{(1)}(\vec{k}) + \zeta^{(3)}(\vec{k}) \, ,
    \end{aligned}
\end{equation}
where $\zeta_G^{[\alpha]}(\vec{q}) = q^{\alpha} \zeta_G(\vec{q})$ and $\zeta^{(n)}$ corresponds to the $n$-th order in $\zeta^{(1)}$ rather than the $n$-th order in LSS perturbation theory. The corresponding primordial power spectrum is
\begin{equation}
    \label{eq:p}
    \begin{aligned}
        \langle \zeta(\vec{k}) \zeta(\vec{k}') \rangle &= \langle \zeta^{(1)}(\vec{k}) \zeta^{(1)}(\vec{k}') \rangle + \mathcal{O}((\zeta^{(1)})^6) \\
        &= (2 \mathbf{\pi})^3 \delta^{(3)}_D(\vec{k}+\vec{k}')P_{\zeta}(k) \, ,
    \end{aligned}
\end{equation}
since $ \langle \zeta^{(1)}(\vec{k}) \zeta^{(3)}(\vec{k}') \rangle =0$ for this specific parity-violating template. 

The trispectrum can be calculated as
\begin{equation}
    \begin{aligned}
        T_{\zeta}(\vec{k}_1,\vec{k}_2,\vec{k}_3, \vec{k}_4) &= (T_{\zeta}^{1113}(\vec{k}_1,\vec{k}_2,\vec{k}_3, \vec{k}_4) + 3 \ \rm{perm.}) + \mathcal{O}((\zeta^{(1)})^8) \, .
    \end{aligned}
\end{equation}
As before, we split the trispectrum into the real (parity-even) and imaginary (parity-odd) parts, $T_{\zeta}= T_{\zeta,+} + i T_{\zeta,-}$. This template does not produce a parity-even part at tree-level\footnote{$T_{\zeta}^{1133}$ is the leading-order parity-even contribution, which is 1-loop.}, whereas the tree-level parity-odd part is given by \cite{coulton_2023}
\begin{equation}
    \label{eq:template}
    T_{\zeta,-}(\vec{k}_1,\vec{k}_2,\vec{k}_3, \vec{k}_4) = - g_- P_{\zeta}(k_1) P_{\zeta}(k_2) P_{\zeta}(k_3) (k_1^{\alpha} \ k_2^{\beta} \ k_3^{\gamma} \ \vec{k}_1 \cdot (\vec{k}_2 \times \vec{k}_3) + 23 \ \rm{perm.}) \, .
\end{equation}
A detailed analysis of this template has been done in Ref.~\cite{Jamieson:2024mau}, where it was shown that the trispectrum peaks in a squeezed limit and reaches its maximum when $k_1 \rightarrow 0 $ with $k_1 \ll k_2 < k_3 < k_4 $. However, it is less sensitive to equilateral configurations where all the $k$-vectors are of similar length, and to folded configurations where the sums of wave vectors go to zero, such as $\vec{k}_1 + \vec{k}_2 \rightarrow 0$.

\subsection{Numerical evaluation of the nonlinear parity-odd trispectrum}
We evaluate the nonlinear parity-odd matter trispectrum from Eq.~\eqref{eq:Teft} for the primordial parity violation provided by the template given in Eq.~\eqref{eq:template}. 

We compute the linear matter power spectrum, $P_m^{11}(k)$, at redshift $z=0$ using the linear Boltzmann solver {\sf CLASS} \cite{Diego_Blas_2011}, with the following cosmological parameters from Ref.~\cite{Planck_2020}: $\Omega_b = 0.0489$,  $\Omega_m = 0.3111$, and $H_0 = 67.66~\mathrm{km~s^{-1}~Mpc^{-1}}$. The primordial power spectrum is given by
\begin{equation}
    P_{\zeta}(k) = \frac{2 \pi^2 A_s}{k^3} \left( \frac{k}{k_p} \right)^{n_s-1} \, ,
\end{equation}
where $A_s = 2.105 \cdot 10^{-9}, ~ n_s = 0.9665$, and $k_p = 0.05 ~ {\rm Mpc}^{-1}$ \cite{Planck_2020}. We obtain the transfer function using the relation $\mathcal{M}^2(k) = P_m^{11}(k)/P_{\zeta}(k)$. In the template given in Eq.~\eqref{eq:template}, we use $\alpha = 0$, $\beta = -1$, $\gamma = -2$, and $g_- = 2 \ (3/5)^2 10^7$.

For $T_{m,-}^{1113,I}$ given in Eq.~\eqref{eq:T1113I}, we use the expression of $P_m^{13}$ given in Eq.~\eqref{eq:p13}, which can be reduced to a one-dimensional integral \cite{Makino:1991rp}. For $T_{m,-}^{1113,II}$ and $T_{m,-}^{1122,I}$ from Eqs.~\eqref{eq:T1113II} and \eqref{eq:T1122I}, we use Monte-Carlo integration to evaluate the three-dimensional integrals. We evaluate the integrals in Cartesian coordinates with $10^7$ evaluations. We take $\Lambda = 20~h~{\rm Mpc}^{-1}$ for the cutoff scale (see Eq.~\eqref{eq:Teft}), where $H_0=100~h~\mathrm{km~s^{-1}~Mpc^{-1}}$.
We exclude a sphere of radius $\epsilon = 10^{-6} ~ h~{\rm Mpc}^{-1}$ around each pole to avoid numerical errors coming from large integrand values, which cancel in the full trispectrum. We have verified that the full trispectrum result is converged with respect to varying $\epsilon$.

\begin{figure}
    \centering
    \includegraphics[width=\linewidth]{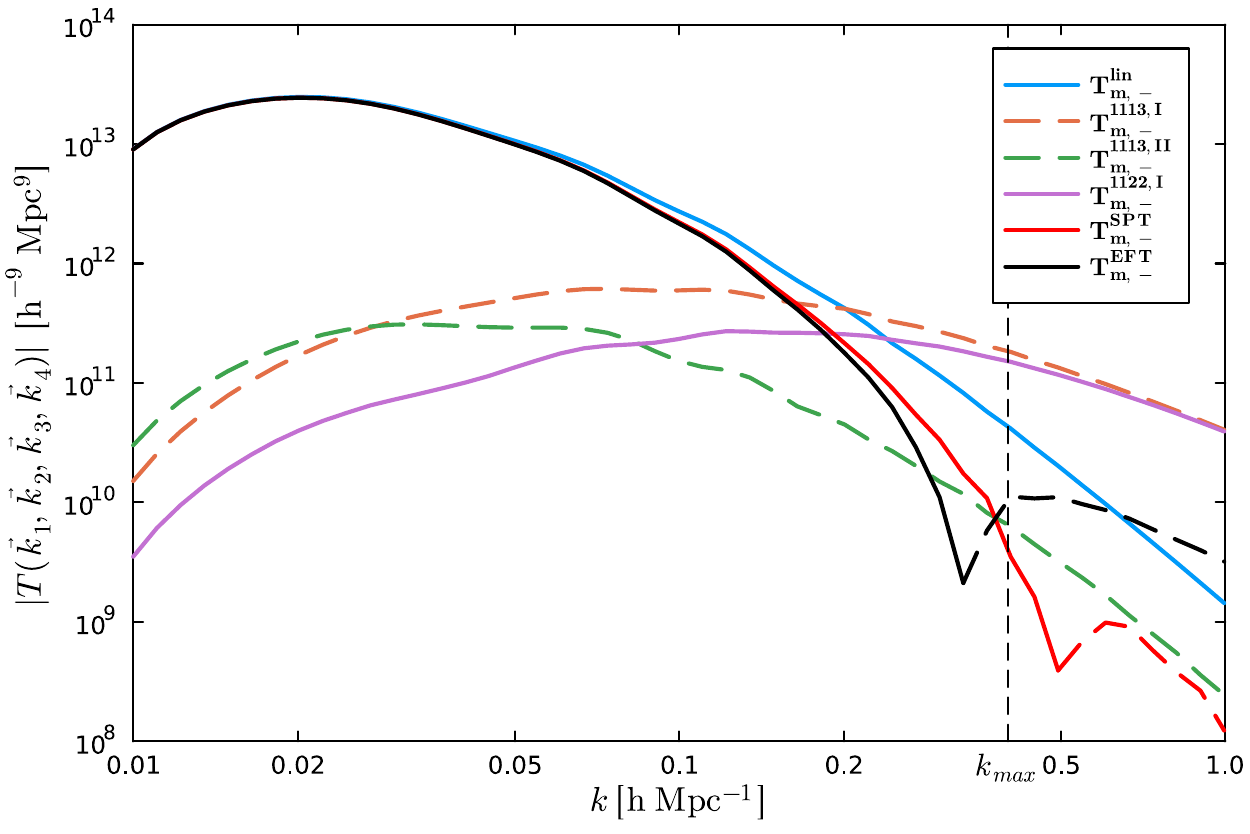}
    \caption{Absolute values of the linear and nonlinear parity-odd trispectra evaluated with the primordial template in Eq.~\eqref{eq:template} in the configuration of Eq.~\eqref{eq:conf} at $z=0.5$. Dashed (solid) lines denote negative (positive) values. The linear trispectrum, $T_{m,-}^{lin} \equiv T_{m,-}^{1111}$, defined in Eq.~\eqref{eq:Tlin}, is shown in blue. We also show the SPT trispectrum from Eq.~\eqref{eq:trispectrum} in red and the EFT trispectrum from Eq.~\eqref{eq:Teft} in black. Nonlinear 1-loop corrections are given by $T_{m,-}^{1113,I}$  (Eq.~\eqref{eq:T1113I}, orange), $T_{m,-}^{1113,II}$ (Eq.~\eqref{eq:T1113II}, green), and $T_{m,-}^{1122,I}$ (Eq.~\eqref{eq:T1122I}, purple), summing all permutations in $(a,b,c,d)$ for $T_{m,-}^{abcd}$. The EFT trispectrum remains reliable up to $k_{max} \simeq 0.4 ~ h~\mathrm{Mpc}^{-1}$. }
    \label{fig:Tlog}
\end{figure}

Figure~\ref{fig:Tlog} shows the absolute value of each nonlinear correction together with the linear and nonlinear SPT and EFT trispectra for the following configurations:  
\begin{equation}
    \label{eq:conf}
    \vec{k}_1 = \begin{pmatrix} 0.005~{h~{\rm Mpc}^{-1}} \\ 0 \\ 0\end{pmatrix} , \ \vec{k}_2 = \begin{pmatrix} 0 \\ 0.65k \\ 0\end{pmatrix}, \ \vec{k}_3=\begin{pmatrix} 0 \\ 0 \\ k\end{pmatrix}, \ \vec{k}_4 = -\vec{k}_3 - \vec{k}_2 - \vec{k}_1 \, .
\end{equation}
This choice of $k$-vectors is particularly illustrative, since $k_1 \ll k_2 < k_3 < k_4$ when $k \gg 0.0077~h~{\rm Mpc}^{-1}$ and, as mentioned in the last section, the template given in Eq.~\eqref{eq:template} peaks for such squeezed configuration.

To evaluate the EFT trispectrum, we use 
the effective sound speed, $c_{s,\Lambda}^2(z)$, found by fitting the 1-loop nonlinear power spectrum to the halofit output from Ref.~\cite{Takahashi_2012, Smith2003StableClustering} at $z=0.5$. The fit is valid up to a scale $k_{max} \simeq 0.4 ~ h ~\mathrm{Mpc}^{-1}$ at the $1 \%$  level. 

In Figure~\ref{fig:Tlog}, we find that for the configurations from Eq.~\eqref{eq:conf}, the nonlinearities dampen the trispectrum at short scales. As depicted through dashed and solid lines, the linear trispectrum and $T_{m,-}^{1122,I}$ are positive, while the $T_{m,-}^{1113,I}$ and $T_{m,-}^{1113,II}$ contributions are negative. $T_{m,-}^{1113}$ (the sum $T_{m,-}^{1113,I}+T_{m,-}^{1113,II}$) dominates $T_{m,-}^{1122,I}$ in magnitude at all scales for this particular primordial trispectrum. This explains the decrease in the amplitude of the SPT trispectrum. The EFT trispectrum suggests that at $z=0.5$ SPT remains reliable up to $k \approx 0.16 ~ h ~ \mathrm{Mpc}^{-1}$ at the $10 \%$ level and $k \approx 0.13 ~ h ~ \mathrm{Mpc}^{-1}$ at the $5 \%$ level.

\subsection{BAO in the parity-odd trispectrum}
As discussed in Ref.~\cite{Hou_2024}, if the detected signal of parity violation has a primordial origin,  BAO must be imprinted on it. Hence, the detection of BAO in the parity-odd trispectrum would make the overall detection more robust and confirm it as a signature of new early-universe physics. To obtain the best constraints from model-dependent searches, a theoretical model is essential. In this section, we illustrate the BAO in our parity-odd trispectrum with the same primordial template from Eq.~\eqref{eq:template}, and show how the nonlinearities alter it.

To demonstrate the effect of nonlinearities on BAO, we construct the so-called ``no-wiggle'' power spectrum as follows. We first Fourier transform the linear matter power spectrum to obtain the two-point correlation function. We remove the BAO peak and smoothly interpolate the correlation function in its place. Transforming this smoothed correlation function back to Fourier space, we obtain the no-wiggle power spectrum, $P^{lin}_{nw}(k)$, which has no BAO feature. This is the procedure from Ref.~\cite{Hou_2024} and is similar to the procedure in Ref.~\cite{Vlah_2015}.

\begin{figure}
    \centering 
    \includegraphics[width=0.65\linewidth]{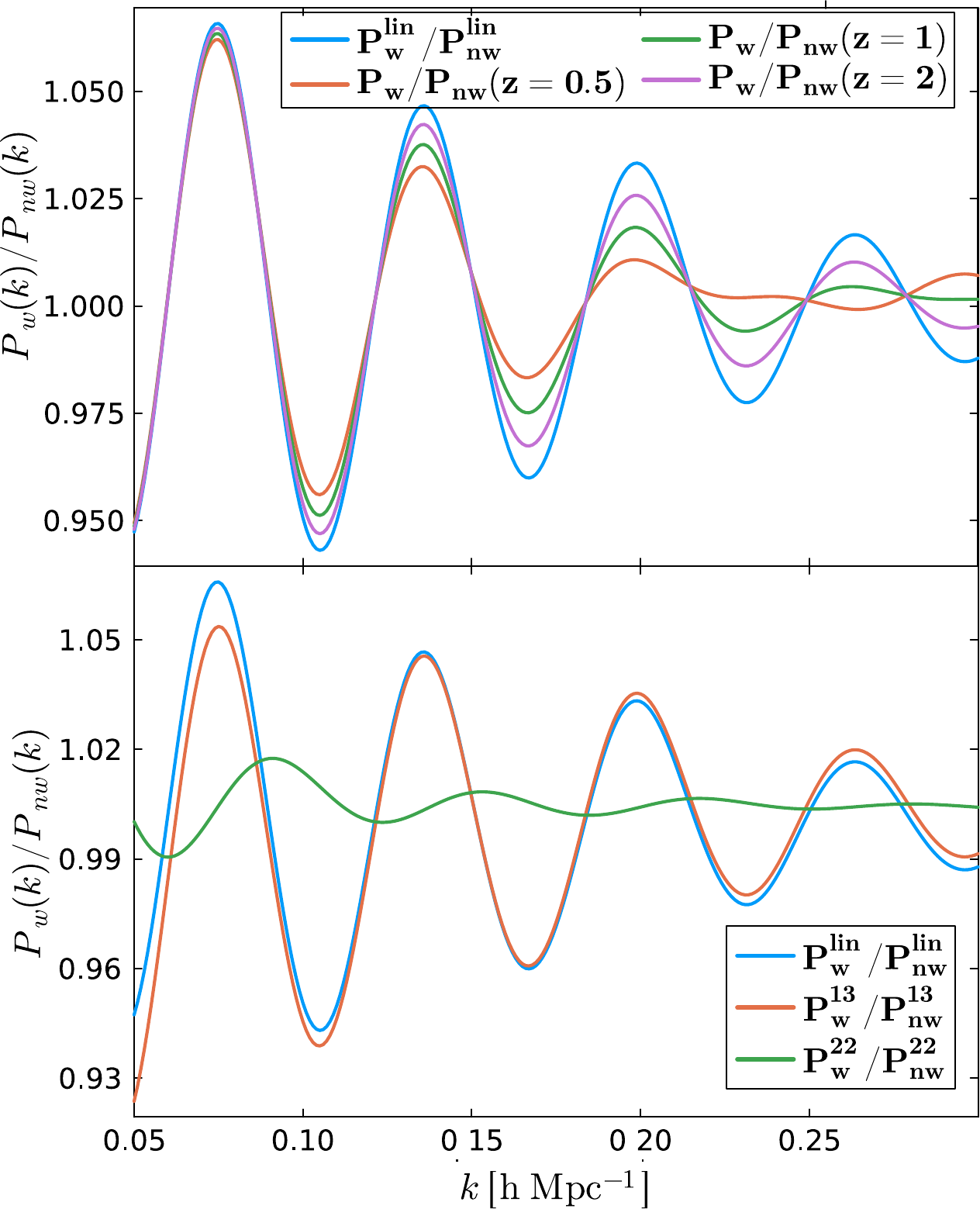}
    \caption{Impact of nonlinearities on the BAO in the power spectrum. Top: Ratios of wiggle to no-wiggle power spectrum for linear (blue) and SPT 1-loop
    at redshifts $z=0.5$ (orange), 1 (green) and 2 (purple). Bottom: Ratios of wiggle to no-wiggle power spectra for linear (blue) and SPT 1-loop corrections, with $P_m^{13}$ in orange and $P_m^{22}$ in green.} 
    \label{fig:PBAO}
\end{figure}

We denote the linear matter power spectrum from {\sf CLASS} with BAO by the wiggle power spectrum, $P^{lin}_w(k)$. We then evaluate the nonlinear 1-loop corrections separately for the no-wiggle and wiggle power spectra and take ratios as depicted in Figure~\ref{fig:PBAO}. The EFT result is similar, so we restrict the discussion to SPT. We reproduce the well-known BAO damping due to nonlinearities~\cite{Eisenstein:2006nj}.

Next, we compute the linear no-wiggle parity-odd matter trispectrum as
\begin{equation}
    \label{eq:Tlin_nw}
    T^{lin}_{nw}(\vec{k}_1, \vec{k}_2, \vec{k}_3, \vec{k}_4) =\mathcal{M}_{nw}(k_1)\mathcal{M}_{nw}(k_2)\mathcal{M}_{nw}(k_3)\mathcal{M}_{nw}(k_4) T_{\zeta,-}(\vec{k}_1, \vec{k}_2, \vec{k}_3, \vec{k}_4) \, ,
\end{equation}
where $\mathcal{M}_{nw}(k)$ is the no-wiggle transfer function given by 
\begin{equation}
   \mathcal{M}_{nw}(k) = \sqrt{\frac{P^{lin}_{nw}(k)}{P_{\zeta}(k)}}\, ,
\end{equation}
and $T_{\zeta,-}$ is the primordial template from Eq.~\eqref{eq:template}. 

\begin{figure}
    \centering
    \includegraphics[width=\linewidth]{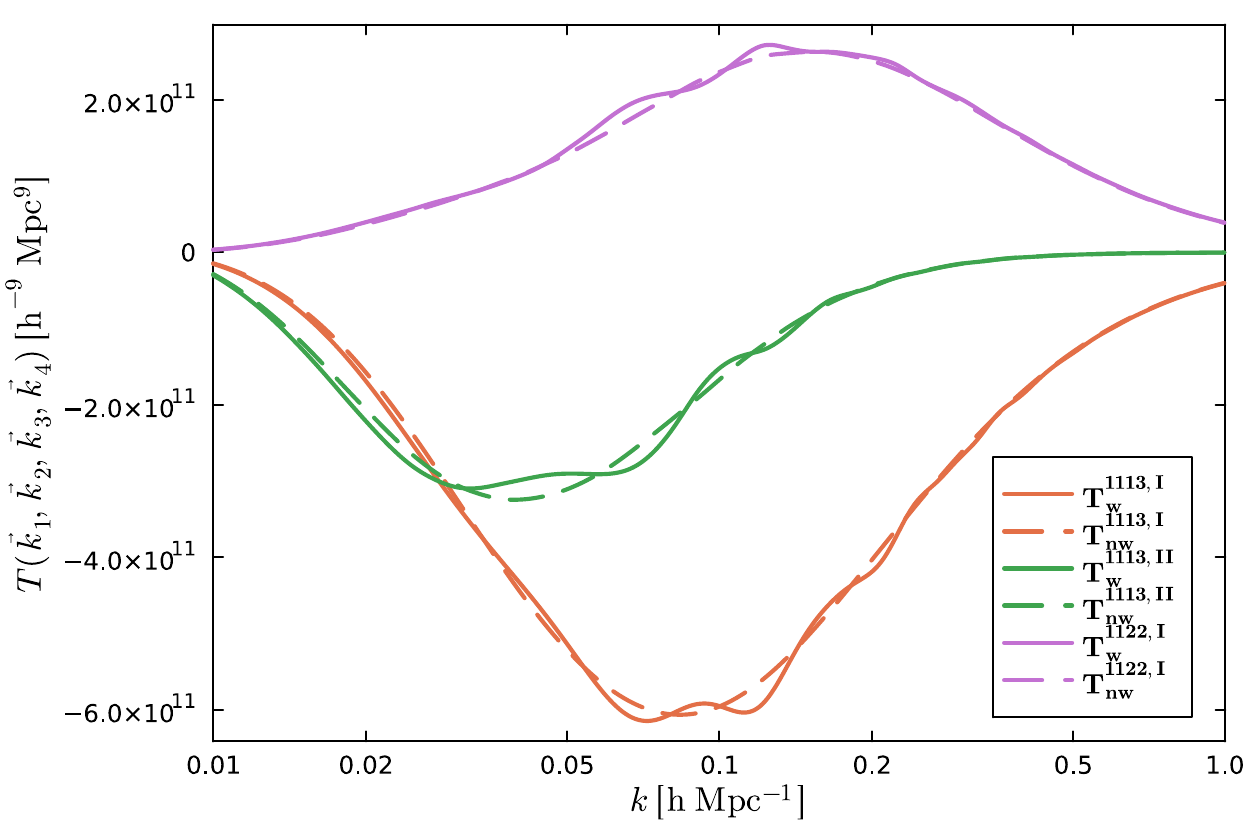}
    \caption{1-loop corrections to the parity-odd trispectrum with the primordial template from Eq.~\eqref{eq:template}, evaluated in the configurations from Eq.~\eqref{eq:conf} at $z=0.5$. The solid lines contain BAO, while the dashed lines do not.
    }
    \label{fig:Tnl}
\end{figure}

We compute the nonlinear no-wiggle trispectrum, $T_{nw}$, by plugging Eq.~\eqref{eq:Tlin_nw} into Eq.~\eqref{eq:trispectrum}. The wiggle and no-wiggle nonlinear trispectrum corrections are depicted in Figure~\ref{fig:Tnl}. We denote our linear parity-odd trispectrum as $T^{lin}_w$ and the nonlinear one as $T_w$, dropping the subscript ``${m,-}$'' in the figure (same for the nonlinear corrections $T^{abcd}_{m,-} \equiv T^{abcd}_{w}$). 

\begin{figure}
    \centering 
    \includegraphics[width=0.65\linewidth]{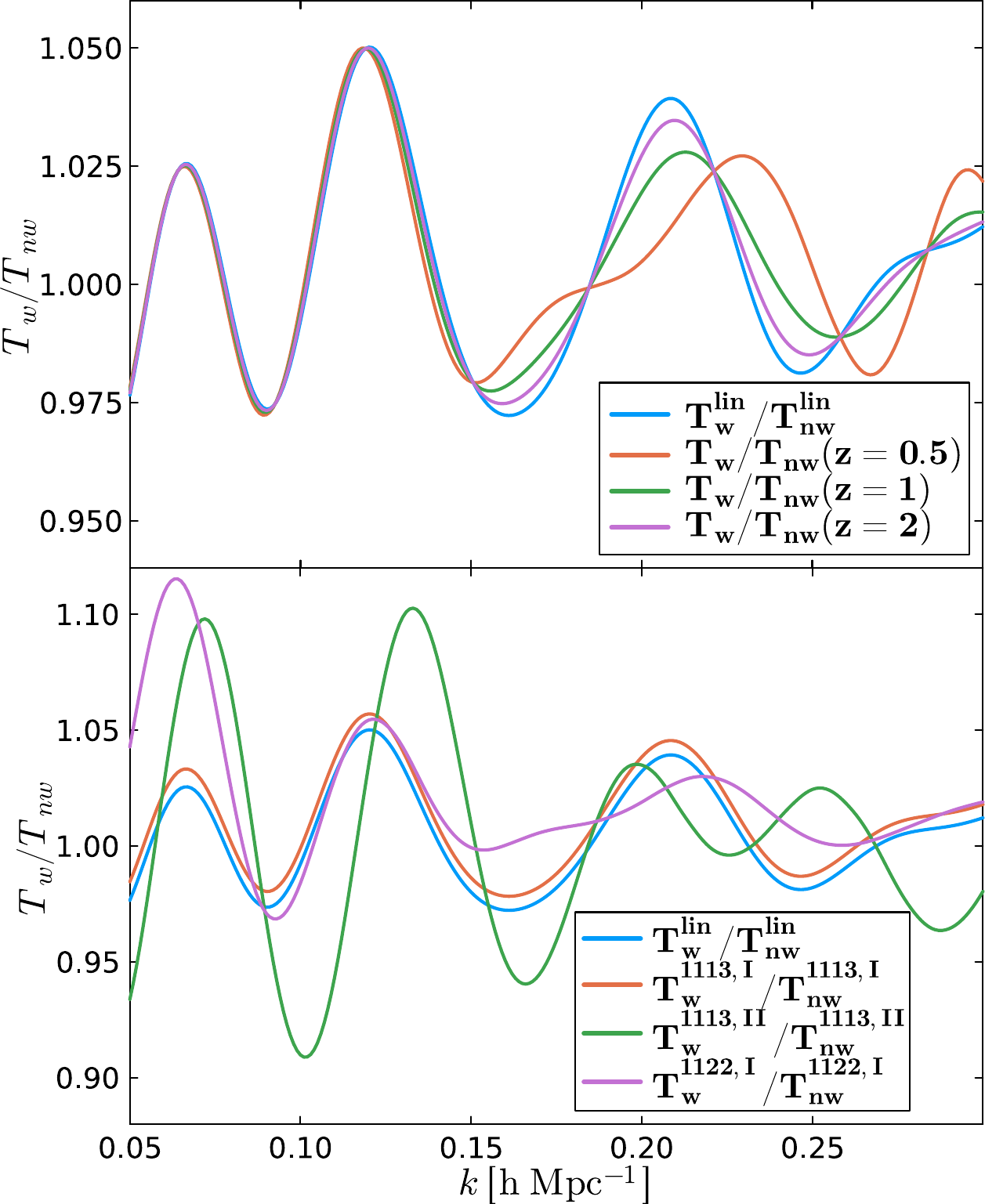}
    \caption{Impact of nonlinearities on the BAO in the parity-odd trispectrum. Top: Ratios of wiggle to no-wiggle trispectrum for linear (blue) and SPT trispectrum at 1-loop at redshifts 0.5 (orange), 1 (green) and 2 (purple). Bottom: Ratios of wiggle to no-wiggle trispectrum for linear (blue) and SPT 1-loop corrections, with $T^{1113,I}$ in orange, $T_{m,-}^{1113,II}$ in green and $T_{m,-}^{1122,I}$ in purple.}
    \label{fig:TBAO}
\end{figure}

We compute the ratios of the wiggle and no-wiggle trispectra for different redshifts as illustrated in Figure~\ref{fig:TBAO}. The top panels in Figures~\ref{fig:PBAO} and \ref{fig:TBAO} show two different damping effects at high-$k$ that are well-known from the power spectrum. The first effect is the decrease of the BAO amplitude in linear theory, due to Silk damping and the finiteness of the decoupling epoch \cite{Silk1968, PeeblesYu1970}. The second effect is the BAO damping in the nonlinear theory, due to large-scale bulk flows and nonlinear clustering \cite{Eisenstein:2006nj}. We find that both of these effects are also present in the nonlinear parity-odd trispectrum. 

Another well-known effect for the nonlinear matter power spectrum is the BAO phase shift due to mode coupling \cite{Crocce_2008, Sherwin:2012nh}. This effect is small for the power spectrum; however, it becomes more pronounced in higher-order statistics, as can be easily seen for the trispectrum in the top panel of Figure~\ref{fig:TBAO}. 

To illustrate the mode couplings, we show the wiggle to no-wiggle ratios of the nonlinear corrections in the bottom panels of Figures~\ref{fig:PBAO} and \ref{fig:TBAO}. In the case of the power spectrum, we see that the oscillations of $P_w^{22}$ are out of phase with those of the linear power spectrum. As shown in Ref.~\cite{Sherwin:2012nh}, the reason for this is that an overdense region is locally like a positively curved universe, leading to a contraction of the correlation function features and additionally to an enhancement of the growth of local perturbations. The same effect is present in the trispectrum: the oscillations of $T_w^{1113,II}$ are out of phase with those of the linear trispectrum. We find that the BAO phase shift is much more pronounced in the trispectrum than in the power spectrum, because overdense regions are weighted more in higher-order statistics.

\section{Conclusions}
\label{sec:conclusions}
The trispectrum of cosmic density perturbations is a powerful probe of cosmological parity violation in the primordial universe. However, nonlinear structure formation modifies the form of the matter trispectrum from its primordial one. In this work, we computed the nonlinear parity-odd matter trispectrum at 1-loop order with a primordial source of parity violation. We introduced our diagrammatic rules that represent correlation functions with PNG in an illuminating and compact way. As gravitational evolution does not source any parity violation, the nonlinear corrections to the parity-odd matter trispectrum can be expressed in terms of the primordial trispectrum. Our analytical result is general and can be applied to any primordial parity-odd trispectrum or different parity-odd templates. 

We showed that the IR modes do not affect the nonlinear corrections to the matter trispectrum, as required by the equivalence principle. We derived the form of the UV counterterm that accounts for the effect of small scales (high-$k$ modes) on large scales. We found that the counterterm has the same scaling as the power spectrum.

By applying our results to a parity-violating primordial template, we demonstrated how the nonlinearities change the linear matter trispectrum at small scales in a squeezed limit. We also illustrated BAO damping and phase shift in the nonlinear trispectrum, analogous to those in the power spectrum, although a more accurate modeling requires an IR resummation \cite{Senatore_2015}, which we leave for future work.

Our work represents the first step towards an accurate modeling of the nonlinear evolution of parity-violating primordial trispectra of cosmic density fields. However, our work is limited to the trispectrum of the underlying matter distribution. In future works, we will extend this analysis to biased tracers for a comparison with observational data.

\acknowledgments
We thank Angelo Caravano, Şafak Çelik, Patricia Diego Palazuelos, Adriaan J. Duivenvoorden, Jiamin Hou, Ippei Obata and Fabian Schmidt for insightful discussions. We also thank the organizers and participants of \textit{Parity Violation from Home 2024} for useful discussions and feedback. This work was supported in part by JSPS KAKENHI Grant Nos.~JP20H05850 and JP20H05859, and the Deutsche Forschungsgemeinschaft (DFG, German Research Foundation) under Germany's Excellence Strategy---EXC-2094---390783311. This work has also received funding from the European Union's Horizon 2020 research and innovation programme under the Marie Skłodowska-Curie Grant Agreement No.~101007633. The Kavli IPMU is supported by World Premier International Research Center Initiative (WPI), MEXT, Japan.
\appendix
\section{SPT kernels}
Solving the Newtonian fluid Eqs.~\eqref{eq:continuity}-\eqref{eq:poisson} gives us the velocity divergence field, $\theta = \vec{\nabla} \cdot \vec{v}$, at linear order as
\begin{equation}
    \vec{\theta}^{(1)}(\vec{k}, \tau) = - \left[\delta_m^{(1)}(\vec{k}, \tau)\right]' = -\mathcal{H}(\tau) f(\tau) \delta_m^{(1)}(\vec{k},\tau) \, ,
\end{equation}
where $f(\tau) = d\ln D/d\ln a$ is the logarithmic growth rate. To solve the nonlinear equations, we expand $\delta_m$ and $\theta$ perturbatively as given in Eq.~\eqref{eq:nonlinmatter} and analogously for $\theta$,
\begin{equation}
    \theta^{(n)}(\vec{k}, \tau) = -\mathcal{H}(\tau) f(\tau) D^n(\tau)\int_{\vec{k}_1,\dots,\vec{k}_n} (2 \mathbf{\pi})^3 \delta^{(3)}_D(\vec{k}-\vec{k}_{12\dots n}) G_n^{(s)}(\vec{k}_1,\dots,\vec{k}_n) \delta^{(1)}(\vec{k}_1)\dots\delta^{(1)}(\vec{k}_n) \, .
\end{equation}
Starting with $F_1^{(s)}(\vec{k}_1) = 1$ and $G_1^{(s)}(\vec{k}_1) = 1$ and solving recursively for the higher orders gives us the symmetric SPT kernels\footnote{The $G_3^{(s)}$ kernel is not required for this work.} given by
\label{app:kernels}
\begin{equation}
    \begin{aligned}
        F_2^{(s)}(\vec{k}_1, \vec{k}_2) &= \frac{5}{7} + \frac{2}{7}  \frac{(\vec{k}_1 \cdot \vec{k}_2)^2}{k_1^2 k_2^2} + \frac{\vec{k}_1 \cdot \vec{k}_2}{2} \left(\frac{1}{k_1^2} + \frac{1}{k_2^2}\right)  \, ,\\ 
        G_2^{(s)}(\vec{k}_1, \vec{k}_2) &= \frac{3}{7} + \frac{4}{7}  \frac{(\vec{k}_1 \cdot \vec{k}_2)^2}{k_1^2 k_2^2} + \frac{\vec{k}_1 \cdot \vec{k}_2}{2} \left(\frac{1}{k_1^2} + \frac{1}{k_2^2}\right)  \, ,\\ 
        F_3^{(s)}(\vec{k}_1, \vec{k}_2, \vec{k}_3) &= \frac{2k^2}{54} \left[ \frac{\vec{k}_1 \cdot \vec{k}_{23}}{k_1^2 k_{23}^2} G_2^{(s)}(\vec{k}_2, \vec{k}_3)  + (2 \ \rm{cyclic}) \right] \\
        &+ \frac{7}{54} \vec{k} \cdot \left[ \frac{\vec{k}_{12}}{k_{12}^2} G_2^{(s)}(\vec{k}_1, \vec{k}_2) + (2 \ \rm{cyclic}) \right] \\
        &+ \frac{7}{54} \vec{k} \cdot \left[ \frac{\vec{k}_1}{k_1^2} F_2^{(s)}(\vec{k}_1, \vec{k}_2) + (2 \ \rm{cyclic}) \right] \, .
    \end{aligned}
\end{equation}
The kernels have the property that \cite{Scoccimarro:1995if}
\begin{equation}
    \begin{aligned}
        &\lim_{q\rightarrow\infty} F_n^{(s)}(\vec{k}_1, \dots,\vec{k}_{n-2}, \vec{q}, -\vec{q}) \propto \frac{k^2}{q^2} \, , \\
        &\lim_{q\rightarrow\infty} F_2^{(s)}(\vec{q}, \vec{k}-\vec{q}) \propto \lim_{q\rightarrow\infty} F_2^{(s)}(\vec{k}_1 + \vec{q}, \vec{k}_2-\vec{q}) \propto \frac{k^2}{q^2} \, ,\\
        &\lim_{q\rightarrow\infty} F_3^{(s)}(-\vec{q}, \vec{k}_1+\vec{q}, \vec{k}_2) \propto\frac{k^2}{q^2} \, ,
    \end{aligned}
\end{equation}
where $k_1 \sim k_2 \sim k$ are of the same order of magnitude.

\section{Primordial five-point correlation function}
\label{app:5pcf}
We assume PNG in the curvature perturbations of the form
\begin{equation}
    \begin{aligned}
        \zeta (\vec{k}) \equiv \sum_{n=1}^{\infty} (\zeta^{(n)}(\vec{k}) -\langle\zeta^{(n)} \rangle) \, ,
    \end{aligned}
\end{equation}
where $\zeta^{(1)}(\vec{k})= \zeta_G(\vec{k})$ is a Gaussian variable and
$\zeta^{(n)}$ can be expanded in terms of $\zeta_G$ as
\begin{equation}
    \zeta^{(n)}(\vec{k}) = \sum_{i=1}^n\int_{\vec{q}_1, \dots,\vec{q}_n} (2\pi)^3 \delta_D(\vec{k} - \vec{q}_1 - \dots - \vec{q}_n) V_n(\vec{q}_1, \dots,\vec{q}_n) \zeta_G(\vec{q}_1) \dots\zeta_G(\vec{q}_n) \, ,
\end{equation}
with $V_n(\vec{q}_1,\dots\vec{q}_n)$ being the vertex factor at $n$-th order in perturbation theory and $V_1(\vec{k)} \equiv 1$. 

 The primordial template we considered in this work, defined in Eq.~\eqref{eq:PNG}, does not produce a primordial five-point function and hence the $T^{1112}_{m,-}$ term vanishes. A possible template that gives rise to a parity-odd five-point function is 
\begin{equation}
    \zeta (\vec{x}) = \zeta_G(\vec{x}) + f_{NL} \zeta^2_G(\vec{x})^2  + g_{NL} \zeta^3_G(\vec{x}) + g_- \vec{\nabla} \zeta_G^{[\alpha]} (\vec{x})\cdot \left[\vec{\nabla} \zeta_G^{[\beta]}(\vec{x}) \times \nabla \zeta_G^{[\gamma]} (\vec{x})\right] \, .
\end{equation}
In this case, the vertex factors are given by 
\begin{equation}
    \begin{aligned}
        V_2(\vec{k}_1, \vec{k}_2) &= f_{NL} \, ,\\
        V_3(\vec{k}_1, \vec{k}_2, \vec{k}_3) &= g_{NL} + g_- k_1^{\alpha} k_2^{\beta} k_3^{\gamma} \ \vec{k}_1 \cdot ( \vec{k}_2 \times \vec{k}_3)\\
        &\equiv V_3^+(\vec{k}_1, \vec{k}_2, \vec{k}_3) + V_3^-(\vec{k}_1, \vec{k}_2, \vec{k}_3) \, .
    \end{aligned}
\end{equation}
The parity-odd five-point function is then
\begin{equation}
    \begin{aligned}
        Q^{11123}_{\zeta,-}(\vec{k}_1, \vec{k}_2, \vec{k}_3, \vec{k}_4, \vec{k}_5) = & \, 6 \, V_2(-\vec{k}_3,\vec{k}_3 + \vec{k}_4) V_3^-(-\vec{k}_1,-\vec{k}_2, \vec{k}_1 + \vec{k}_2 + \vec{k}_5) \\
        &P_{\zeta}(k_1) P_{\zeta}(k_2) P_{\zeta}(k_3) P_{\zeta}(|\vec{k}_3 + \vec{k}_4|) + (k_3 \leftrightarrow k_1) + (k_3 \leftrightarrow k_2) \, ,
    \end{aligned}
\end{equation}
so that the full parity-odd five-point function at leading order is
\begin{equation}
    Q_{\zeta,-}(\vec{k}_1, \vec{k}_2, \vec{k}_3, \vec{k}_4, \vec{k}_5) = Q^{11123}_{\zeta,-}(\vec{k}_1, \vec{k}_2, \vec{k}_3, \vec{k}_4, \vec{k}_5) + 19 \ \rm{perm.} \, .
\end{equation}

\section{Details on the IR limit}
\label{app:IR}
In this section, we give more details on the derivation of the IR limits ($q \ll k_i$) of $T^{1122,I}_{m,-}$ and $T^{1113,II}_{m,-}$ from Eqs.~\eqref{eq:1122IR} and \eqref{eq:1113II_IR}, respectively.

We start with $T_{m,-}^{1122,I}$. We find
\begin{equation}
    \begin{aligned}
        T_{m,-}^{1122,I}(\vec{k}_1, \vec{k}_2, &\vec{k}_3, \vec{k}_4)\Big{|}_{IR} =  4 \int \frac{d \Omega }{(2 \pi)^3}\\
        &\int_{q \ll k} dq q^2 F_2^{(s)}(\vec{q}, \vec{k}_3 -\vec{q}) \ F_2^{(s)}(-\vec{q}, \vec{k}_4 +\vec{q}) \ P_m^{11}(q) T_{m,-}^{1111}(\vec{k}_1, \vec{k}_2, \vec{k}_3 - \vec{q}, \vec{k}_4 + \vec{q}) \\
        &= 4 ~ T_{m,-}^{1111}(\vec{k}_1, \vec{k}_2, \vec{k}_3, \vec{k}_4)\int \frac{d \Omega }{(2 \pi)^3} \frac{\left[7 k_3^3 (\hat{q} \cdot \hat{k}_3)\right]\left[- 7 k_4^3 (\hat{q} \cdot \hat{k}_4)\right]}{196 k_3^2 k_4^2} \int_{q \ll k} dq ~ P_m^{11}(q)  \\
        &= - \frac{1}{3 (2 \pi)^3} k_3 k_4 T_{m,-}^{1111}(\vec{k}_1, \vec{k}_2, \vec{k}_3, \vec{k}_4) \int d\Omega (\hat{q} \cdot \hat{k}_3) (\hat{q} \cdot \hat{k}_4) \int_{q \ll k} dq ~ P_m^{11}(q)\\
        & \equiv - \frac{1}{3(2 \pi)^3} k_3 k_4 T_{m,-}^{1111}(\vec{k}_1, \vec{k}_2, \vec{k}_3, \vec{k}_4) I_{\Omega}(\hat{k}_3 \cdot \hat{k}_4) \int_{q \ll k} dq ~P_m^{11}(q) \, ,
    \end{aligned}
    \label{eq:1122Iapp}
\end{equation}
where we have defined $d\Omega = d\phi d\cos\theta$. Using the first Legendre polynomial, $P_1(x) = x$, and the addition theorem for spherical harmonics,
\begin{equation}
    P_l(\vec{x} \cdot \vec{y}) = \frac{4 \pi}{2l+1} \sum_{m=-l}^{l} Y_{lm}^*(\vec{x}) Y_{lm}(\vec{y}) \, ,
\end{equation}
we can evaluate the angular integral as
\begin{equation}
    \begin{aligned}
        I_{\Omega}(\hat{k}_3 \cdot \hat{k}_4) &= \int d\Omega(\hat{q} \cdot \hat{k}_3) (\hat{q} \cdot \hat{k}_4) \\
        &= \int d\Omega P_1(\hat{k}_3  \cdot \hat{q}) P_1(\hat{q} \cdot \hat{k}_4) \\
        &= \int d\Omega \left[\frac{4 \pi}{3} \sum_{m=-1}^{1} Y_{1m}^*(\hat{q}) Y_{1m}(\hat{k}_3) \right] \left[\frac{4 \pi}{3} \sum_{m'=-1}^{1} Y_{1m'}^*(\vec{k}_4) Y_{1m'}(\hat{q}) \right] \\
        &= \left(\frac{4 \pi}{3} \right)^2 \sum_{m=-1}^{1} \sum_{m'=-1}^{1} Y_{1m'}^*(\hat{k}_4) Y_{1m}(\hat{k}_3) \int d\Omega~Y_{1m}^*(\hat{q}) Y_{1m'}(\hat{q}) \, .
    \end{aligned} 
\end{equation}
Using the orthonormality property of spherical harmonics,
\begin{equation}
    \int d\Omega~ Y_{lm}^*(\theta, \phi) Y_{l'm'}(\theta, \phi) = \delta_{ll'} \delta_{mm'} \, ,
\end{equation}
we obtain
\begin{equation}
    \begin{aligned}
        I_{\Omega}(\hat{k}_3 \cdot \hat{k}_4) &= \left(\frac{4 \pi}{3} \right)^2 \sum_{m=-1}^{1} Y_{1m}^*(\hat{k}_4) Y_{1m}(\hat{k}_3) \\
        &= \frac{4 \pi}{3} P_1(\hat{k}_4 \cdot \hat{k}_3) = \frac{4 \pi}{3} \hat{k}_4 \cdot \hat{k}_3 \, .
    \end{aligned}
\end{equation}
Hence, Eq.~\eqref{eq:1122Iapp} reduces to
\begin{equation}
    T_{m,-}^{1122,I}(\vec{k}_1, \vec{k}_2, \vec{k}_3, \vec{k}_4)\Big{|}_{IR} = - \frac{2 }{3 (2\pi)^2} T_{m,-}^{1111}(\vec{k}_1, \vec{k}_2, \vec{k}_3, \vec{k}_4) \vec{k}_4 \cdot \vec{k}_3 \int_{q \ll k} dq ~ P_m^{11}(q)\, .
\end{equation}
Taking all the permutations and using $\vec{k}_{1234}= 0$, which means 
\begin{equation}
    \begin{aligned}
        -k_1^2 - k_2^2 -k_3^2 -k_4^2 = 2 (\vec{k}_1 \cdot \vec{k}_2 + \vec{k}_1 \cdot \vec{k}_3 + \vec{k}_1 \cdot \vec{k}_4 + \vec{k}_2 \cdot \vec{k}_3 + \vec{k}_2 \cdot \vec{k}_4 + \vec{k}_3 \cdot \vec{k}_4) \, ,
    \end{aligned}
\end{equation}
we arrive at the final result 
\begin{equation}
    \begin{aligned}
        (T_{m,-}^{1122,I}(\vec{k}_1, \vec{k}_2, \vec{k}_3, \vec{k}_4) &+ 5 \ \rm{perm.})\Big{|}_{IR} = \\
        &\frac{1}{3 (2\pi)^2} T_{m,-}^{1111}(\vec{k}_1, \vec{k}_2, \vec{k}_3, \vec{k}_4) (k_1^2 + k_2^2 + k_3^2 + k_4^2) \int_{q \ll k} dq ~ P_m^{11}(q) \, .
    \end{aligned}
\end{equation}

Next, we show that the IR limit of $T_{m,-}^{1113,II}$ is zero. 

We expand the $F_3^{(s)}$ kernel around $q=0$ and find
\begin{equation}
    \begin{aligned}
        T_{m,-}^{1113,II}(\vec{k}_1, &\vec{k}_2, \vec{k}_3, \vec{k}_4)\Big{|}_{IR} \simeq 3 P_m^{(11)}(k_1) T_{m,-}^{1111}(\vec{k}_1, \vec{k}_2, \vec{k}_3, \vec{k}_4) \\
        &\int \frac{d\Omega}{(2\pi)^3} \int_{q \ll k} dq q^2 \left[f_1(\vec{k}_1, \vec{k}_4) \frac{1}{q} + f_2(\vec{k}_1, \vec{k}_4) + f_3(\vec{k}_1, \vec{k}_4) q + \mathcal{O}(q^2) \right]  \\
        &+ (\vec{k}_1 \leftrightarrow \vec{k}_2 )+ (\vec{k}_1 \leftrightarrow \vec{k}_3) \, ,
    \end{aligned}
\end{equation}
where $f_1, \, f_2$ and $f_3$ are some scalar functions of $\vec{k}_1$ and $\vec{k}_4$.
Since there is no term in the expansion that scales as $q^{-2}$, the IR limit vanishes.

\bibliographystyle{JHEP}
\bibliography{references}
\end{document}